\documentclass[apj,numberedappendix]{emulateapj-rtx4}
\usepackage{natbib}
\usepackage{comment}
\usepackage{amsmath}

\shorttitle{CALIBRATING EXTINCTION-FREE SFRs in NGC\,6946}
\shortauthors{MURPHY ET AL.}
\slugcomment{Draft version 2.4; Accepted to \apj~May 24, 2011}
\journalinfo{Draft version 2.4; Accepted to \apj~May 24, 2011}

\begin{document}
\title{Calibrating Extinction-Free Star Formation Rate Diagnostics with 33\,GH{\MakeLowercase z} Free-Free Emission in NGC\,6946}

\author{E.J.~Murphy\altaffilmark{1}, J.J.~Condon\altaffilmark{2}, E.~Schinnerer\altaffilmark{3}, R.C.~Kennicutt,~Jr.\altaffilmark{4}, D.~Calzetti\altaffilmark{5}, 
L.~Armus \altaffilmark{1}, G.~Helou\altaffilmark{6}, J.L.~Turner\altaffilmark{7}, G.~Aniano\altaffilmark{8}, P.~Beir\~{a}o\altaffilmark{1}, A.D.~Bolatto\altaffilmark{9},  
B.R.~Brandl\altaffilmark{10}, K.V.~Croxall\altaffilmark{11}, D.A.~Dale\altaffilmark{12}, J.L.~Donovan Meyer\altaffilmark{13}, 
B.T.~Draine\altaffilmark{8}, C.~Engelbracht\altaffilmark{14},  L.K.~Hunt\altaffilmark{15}, C.-N.~Hao\altaffilmark{16}, J.~Koda\altaffilmark{13}, H.~Roussel\altaffilmark{17}, 
R.~Skibba\altaffilmark{14}, and J.-D.T.~Smith\altaffilmark{11}}
\altaffiltext{1}{\scriptsize {\it Spitzer Science Center,} California Institute of Technology, MC 314-6, Pasadena CA, 91125, USA; emurphy@ipac.caltech.edu} 
\altaffiltext{2}{\scriptsize National Radio Astronomy Observatory, 520 Edgemont Road, Charlottesville, VA 22903, USA}
\altaffiltext{3}{\scriptsize Max Planck Institut f\"{u}r Astronomie, K\"{o}nigstuhl 17, Heidelberg D-69117, Germany}
\altaffiltext{4}{\scriptsize Institute of Astronomy, University of Cambridge, Madingley Road, Cambridge CB3 0HA, UK}
\altaffiltext{5}{\scriptsize Department of Astronomy, University of Massachusetts, 710 N. Pleasant Street, Amherst, MA 01003, USA}
\altaffiltext{6}{\scriptsize California Institute of Technology, MC 314-6, Pasadena, CA 91125, USA}
\altaffiltext{7}{\scriptsize Department of Physics and Astronomy, UCLA, Los Angeles, CA 90095, USA}
\altaffiltext{8}{\scriptsize Department of Astrophysical Sciences, Princeton University, Princeton, NJ 08544, USA}
\altaffiltext{9}{\scriptsize Department of Astronomy, University of Maryland, College Park, MD 20742, USA}
\altaffiltext{10}{\scriptsize Leiden Observatory, Leiden University, P.O. Box 9513, 2300 RA Leiden, The Netherlands}
\altaffiltext{11}{\scriptsize Department of Physics and Astronomy, University of Toledo,Toledo, OH 43606, USA}
\altaffiltext{12}{\scriptsize Department of Physics \& Astronomy, University of Wyoming, Laramie, WY 82071, USA    }
\altaffiltext{13}{\scriptsize Department of Physics and Astronomy, SUNY Stony Brook, Stony Brook, NY 11794-3800, USA}
\altaffiltext{14}{\scriptsize Steward Observatory, University of Arizona, Tucson, AZ 85721, USA}
\altaffiltext{15}{\scriptsize INAF - Osservatorio Astrofisico di Arcetri, Largo E. Fermi 5, 50125 Firenze, Italy}
\altaffiltext{16}{\scriptsize Tianjin Astrophysics Center, Tianjin Normal University, Tianjin 300387, China}
\altaffiltext{17}{\scriptsize Institut d'Astrophysique de Paris, UMR7095 CNRS Universit\'e Pierre \& Marie Curie, 98 bis Boulevard Arago, 75014 Paris, France}

\begin{abstract}  
Using free-free emission measured in the Ka-band ($26-40$\,GHz) for 10 star-forming regions in the nearby galaxy NGC\,6946, including its starbursting nucleus, we compare a number of star formation rate (SFR) diagnostics that are typically considered to be unaffected by interstellar extinction.  
These diagnostics include non-thermal radio (i.e., 1.4\,GHz), total infrared (IR; $8-1000\,\mu$m), and warm dust (i.e., 24\,$\mu$m) emission, along with hybrid indicators that attempt to account for obscured and unobscured emission from star-forming regions including H$\alpha + 24\,\mu$m and UV + IR measurements.  
The assumption is made that the 33\,GHz free-free emission provides the most accurate measure of the current SFR.  
Among the extranuclear star-forming regions, the 24\,$\mu$m, H$\alpha+24\,\mu$m and UV + IR SFR calibrations are in good agreement with the 33\,GHz free-free SFRs.  
However, each of the SFR calibrations relying on some form of dust emission overestimate the nuclear SFR by a factor of $\sim$2 relative to the 33\,GHz free-free SFR.  
This is more likely the result of excess dust heating through an accumulation of non-ionizing stars associated with an extended episode of star formation in the nucleus rather than increased competition for ionizing photons by dust.  
SFR calibrations using the non-thermal radio continuum yield values which only agree with the 33\,GHz free-free SFRs for the nucleus, and underestimate the SFRs from the extranuclear star-forming regions by an average factor of $\sim$2 and $\sim$4$-$5 before and after subtracting local background emission, respectively.   
This result likely arises from the CR electrons decaying within the starburst region with negligible escape,   
whereas the transient nature of star formation in the young extranuclear star-forming complexes allows for CR electrons to diffuse significantly further than dust heating photons, resulting in an underestimate of the true SFR.  
Finally, we find that the SFRs estimated using the total 33\,GHz flux density appear to agree well with those from using the free-free emission due to the large thermal fractions present at these frequencies even when local diffuse backgrounds are {\it not} removed.  
Thus, rest-frame 33\,GHz observations may act as a reliable method to measure the SFRs of galaxies at increasingly high redshift without the need of ancillary radio data to account for the non-thermal emission.  
\end{abstract}


\section{Introduction}
Accurate estimates of star formation rates (SFRs) are crucial in nearly all fields of extragalactic astronomy.  
Besides requiring a high level of accuracy, the ease with which such estimates can be made plays a critical role in the overall utility of a SFR diagnostic given that, for studying galaxy evolution, observations need to be made over large areas of sky and be deep enough to detect galaxies at high redshifts.  
Presently, a wide variety of SFR diagnostics are commonly used in the literature \citep[e.g.,][]{rck98}, each of which have their strengths and shortcomings.    

Rest-frame UV ($1250-2500$\AA) and optical (e.g., H$\alpha$) observations provide a measure of the radiant energy released by newly formed massive stars.  
While directly related to the massive star formation process, such observations are hampered by interstellar extinction that varies enormously within and among galaxies.  
This is especially problematic for trying to characterize the evolution of the SFR density with redshift as the galaxies that dominate the stellar mass assembly between $1\la z \la 3$ appear to be dusty starbursts with infrared (IR; $8-1000~\micron$) luminosities ranging between $10^{10}~L_{\odot} \la L_{\rm IR} \la 10^{13}~L_{\sun}$ 
 \citep[e.g.,][]{ce01,tak05, rjb09, ejm11a}.  
In these IR-bright starbursts, more than $\ga$90\% of the UV emission can be obscured \citep[e.g.,][]{jh10,vb10,ejm11a}.  

Rest-frame IR measurements, particularly those in the far-infrared (FIR; $42.5-122.5\,\mu$m), 
do not suffer significantly from extinction as they probe the re-radiated UV/optical emission from heated dust grains surrounding massive star-forming regions.  
While an extinction-free measure, the interpretation of IR emission alone is also complex.  
Variations in the dust composition, content, and distribution along the line of sight will affect the fraction of UV photons absorbed, while a portion of the IR emission will arise from dust heated by older stars \citep[i.e., the ÒcirrusÓ component][]{gxh86,lh87, hira03, gjb10}.  

Radio continuum emission is also widely used as a tracer of star formation in galaxies both at low and high redshift as the result of the tight empirical correlation between their optically thin synchrotron emission at 1.4\,GHz 
and their FIR radiation \citep[e.g.,][]{de85,gxh85}.  
While this relation (i.e., the FIR-radio correlation) is certainly rooted in common dependencies on massive star formation, it is unclear how presumably unrelated physical processes affecting the propagation of cosmic-ray (CR) electrons and the heating of dust grains work together to yield a nearly ubiquitous correlation over nearly 5 orders of magnitude in luminosity \citep{yrc01}.  
{\it Spitzer} revealed local correlations in the spatial distributions of 70\,$\mu$m and non-thermal radio emission in the disks of galaxies that reflect an age effect indicating the complexities of using non-thermal radio continuum emission as a SFR diagnostic \citep{ejm06b,ejm08}.    
Additional complications for using the non-thermal radio emission as a SFR diagnostic are also evident given that the correlation between the IR and radio emission is strikingly different for spiral arms and interarm regions \citep{gd11}.  

Higher frequency radio continuum emission, which becomes dominated by thermal (free-free) radiation once beyond $\sim$30\,GHz for globally integrated measurements of star-forming galaxies \citep[e.g.,][]{jc92}, is both largely extinction free and can be directly related to the ionizing photon rate arising from newly formed massive stars.  
Evidence for this has been indicated by detailed studies of star-forming regions in the Galaxy \citep[e.g.,][]{pgm67a}, nearby dwarf irregulars \citep[e.g.][]{kg86},  the nuclei of normal galaxies \citep[e.g.,][]{th83,th94} and starbursts \citep[e.g.,][]{kwm88,th85}, as well as high resolution investigations of super star clusters within nearby blue compact dwarfs \citep[e.g.,][]{thb98, kj99}.  

While typically considered an ideal measure for the current star formation activity in galaxies, the presence of an `anomalous' dust emission component in excess of the free-free emission between $\sim$10 and 90\,GHz, generally attributed to rapidly rotating ultrasmall grains with a non-zero electric moment \citep[e.g.,][]{dl98b,plsd11}, may complicate this picture.  
For a single outer-disk star-forming region in NGC\,6946, \citet{ejm10} detected excess 33\,GHz emission relative to what is expected given existing lower frequency radio data.  
This result is interpreted as the first likely detection of so-called anomalous dust emission outside of the Milky Way.  
Given that the excess was only detected for a single region, this emission component may be negligible for globally integrated measurements.  

In this paper we quantitatively compare SFR estimates using a number of ``extinction-free" diagnostics for 10 star-forming regions within the nearby galaxy NGC\,6946 presented in \citet{ejm10}. 
This is done to assess their reliability against SFRs measured via free-free emission at 33\,GHz.  
The paper is organized as follows:  
In $\S$2 we introduce the data sets used for our SFR comparisons.  
Then, in $\S$3, we calibrate each of the observed quantities for the same initial mass function (IMF).  
Comparisons between each of the SFR diagnostics are made in $\S$4 and discussed in $\S$5.  
Our main conclusions are summarized in $\S$6.  

\begin{figure}
\begin{center}
\scalebox{1.1}{
\plotone{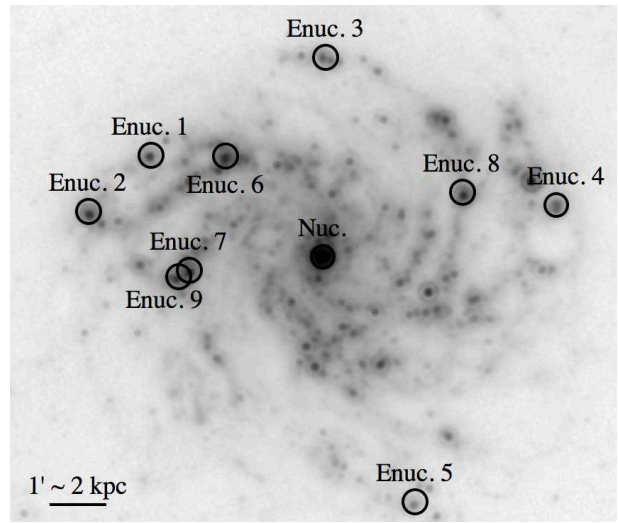}}
\caption{
The 24\,$\mu$m map of NGC\,6946 showing the location of each star-forming region in the present investigation. 
The diameter of each circle corresponds to 25\arcsec, the FWHM of the GBT beam at 33\,GHz, which sets the resolution of this study.  
This diameter projects to a linear scale of 0.8\,kpc at the distance of NGC\,6946.  
\label{fig-1}}
\end{center}
\end{figure}

\begin{deluxetable*}{ccccccccc}
\tablecaption{Source Positions and Radio Photometry \label{tbl-1}}
\tabletypesize{\scriptsize} 
\tablewidth{0pt}
\tablehead{
\colhead{ID}  & \colhead{R.A.} & \colhead{Decl.} &
\colhead{$S_{\rm 33~GHz}$}  & \colhead{$S_{\rm 8.5~GHz}$}  & \colhead{$S_{\rm 4.9~GHz}$}  & 
\colhead{$S_{\rm 1.7~GHz}$}  & \colhead{$S_{\rm 1.5~GHz}$}  & \colhead{$S_{\rm 1.4~GHz}$}\\
\colhead{} & \colhead{(J2000)} & \colhead{(J2000)} &
\colhead{(mJy)} & \colhead{(mJy)} & \colhead{(mJy)}  &
 \colhead{(mJy)} & \colhead{(mJy)} & \colhead{(mJy)}
}
\startdata
     Nucleus   &20~34~52.34   &+60~ 9~14.2  &15.7$\pm$0.79  &31.8$\pm$1.59  &44.8$\pm$2.24  &77.6$\pm$3.88  &83.2$\pm$1.59  & 85.5$\pm$1.59  \\
     Enuc. 1   &20~35~16.65   &+60~11~ 1.1  & 1.0$^{\dagger}\pm$0.05  & 0.9$\pm$0.12  & 0.9$\pm$0.09  & 1.1$\pm$0.15  & 1.1$\pm$0.12  &  1.3$\pm$0.12  \\
     Enuc. 2   &20~35~25.49   &+60~10~ 1.8  & 2.4$\pm$0.16  & 2.6$\pm$0.17  & 3.1$\pm$0.17  & 4.3$\pm$0.26  & 4.5$\pm$0.17  &  5.0$\pm$0.17  \\
     Enuc. 3   &20~34~51.89   &+60~12~44.8  & 1.0$\pm$0.07  & 0.9$\pm$0.12  & 1.1$\pm$0.10  & 1.6$\pm$0.16  & 1.7$\pm$0.12  &  1.8$\pm$0.12  \\
     Enuc. 4   &20~34~19.17   &+60~10~ 8.7  & 2.9$\pm$0.15  & 1.4$\pm$0.13  & 1.8$\pm$0.12  & 2.0$\pm$0.17  & 2.1$\pm$0.13  &  2.3$\pm$0.13  \\
     Enuc. 5   &20~34~39.27   &+60~ 4~55.1  & 0.6$^{\dagger}\pm$0.05  & 0.6$\pm$0.11  & 0.6$\pm$0.09  & 1.0$\pm$0.15  & 1.0$\pm$0.11  &  1.2$\pm$0.11  \\
     Enuc. 6   &20~35~ 6.09   &+60~11~ 0.6  & 2.8$\pm$0.18  & 3.6$\pm$0.21  & 4.3$\pm$0.23  & 6.7$\pm$0.37  & 7.2$\pm$0.21  &  7.6$\pm$0.21  \\
     Enuc. 7   &20~35~11.21   &+60~ 8~59.7  & 3.1$\pm$0.20  & 2.9$\pm$0.18  & 3.6$\pm$0.20  & 4.8$\pm$0.28  & 5.1$\pm$0.18  &  5.3$\pm$0.18  \\
     Enuc. 8   &20~34~32.52   &+60~10~22.0  & 2.2$^{\dagger}\pm$0.14  & 2.0$\pm$0.15  & 2.4$\pm$0.14  & 2.9$\pm$0.20  & 3.1$\pm$0.15  &  3.5$\pm$0.15  \\
     Enuc. 9   &20~35~12.71   &+60~ 8~52.8  & 2.4$\pm$0.16  & 2.5$\pm$0.17  & 2.9$\pm$0.17  & 4.2$\pm$0.25  & 4.4$\pm$0.17  &  4.9$\pm$0.17  
\enddata
\tablecomments{$^{\dagger}$ Correction (`boosting') factors arising from an over subtraction of signal by reference beams have been applied to 33~GHz flux densities for sources in which the correction was larger than 15\%. These scale factors are 2.55, 1.17, and 1.25 for Enuc. 1, 5, and 8, respectively.}
\end{deluxetable*}

\begin{deluxetable*}{ccccccccc}
\tablecaption{Infrared Photometry \label{tbl-2}}
\tabletypesize{\scriptsize} 
\tablewidth{0pt}
\tablehead{
\colhead{ID}  & 
\colhead{$f_{\nu}$ (8\,$\mu$m)$^{\dagger}$}  & \colhead{$f_{\nu}$ (24\,$\mu$m)}  & \colhead{$f_{\nu}$ (70\,$\mu$m)} &  \colhead{$f_{\nu}$ (100\,$\mu$m)}  & 
\colhead{$f_{\nu}$ (160\,$\mu$m)}  & \colhead{$f_{\nu}$ (250\,$\mu$m)}  & \colhead{$f_{\nu}$ (350\,$\mu$m)}  & \colhead{$f_{\nu}$ (850\,$\mu$m)} \\ 
\colhead{} &
\colhead{(mJy)} & \colhead{(mJy)} & 
\colhead{(Jy)} & \colhead{(Jy)} &  \colhead{(Jy)} & \colhead{(Jy)} & \colhead{(Jy)} & 
\colhead{(mJy)}
}
\startdata
     Nucleus   &1506.7$\pm$  75.4  &4308.9$\pm$ 215.4  & 26.11$\pm$  2.61  & 32.09$\pm$  3.21  & 23.67$\pm$  4.74  &  6.80$\pm$  1.09  &  3.12$\pm$  0.50  & 401.8$\pm$  62.8\\
     Enuc. 1   &  33.6$\pm$   2.8  &  88.0$\pm$   4.5  &  0.85$\pm$  0.13  &  1.23$\pm$  0.18  &  1.02$\pm$  0.23  &  0.33$\pm$  0.06  &  0.15$\pm$  0.04  &  10.8$\pm$  17.9\\
     Enuc. 2   &  53.0$\pm$   3.5  & 175.1$\pm$   8.8  &  1.28$\pm$  0.16  &  1.94$\pm$  0.24  &  1.65$\pm$  0.34  &  0.48$\pm$  0.08  &  0.22$\pm$  0.05  &   \nodata \\
     Enuc. 3   &  40.6$\pm$   3.1  &  66.4$\pm$   3.5  &  1.00$\pm$  0.14  &  1.54$\pm$  0.21  &  1.30$\pm$  0.28  &  0.36$\pm$  0.07  &  0.20$\pm$  0.05  &   \nodata \\
     Enuc. 4   &  51.9$\pm$   3.5  &  83.1$\pm$   4.3  &  1.00$\pm$  0.14  &  1.65$\pm$  0.21  &  1.49$\pm$  0.32  &  0.35$\pm$  0.07  &  0.17$\pm$  0.04  &  15.2$\pm$  17.9\\
     Enuc. 5   &  12.5$\pm$   2.4  &  24.7$\pm$   1.6  &  0.40$\pm$  0.11  &  0.57$\pm$  0.15  &  0.60$\pm$  0.16  &  0.16$\pm$  0.04  &  0.11$\pm$  0.04  &   \nodata \\
     Enuc. 6   & 192.3$\pm$   9.9  & 352.2$\pm$  17.6  &  2.62$\pm$  0.28  &  3.77$\pm$  0.40  &  3.02$\pm$  0.61  &  0.69$\pm$  0.12  &  0.37$\pm$  0.07  &  54.7$\pm$  19.6\\
     Enuc. 7   & 186.6$\pm$   9.6  & 366.0$\pm$  18.3  &  3.21$\pm$  0.34  &  4.37$\pm$  0.46  &  3.65$\pm$  0.74  &  1.21$\pm$  0.20  &  0.57$\pm$  0.10  &  60.5$\pm$  20.0\\
     Enuc. 8   & 114.7$\pm$   6.2  & 178.5$\pm$   9.0  &  1.55$\pm$  0.19  &  2.27$\pm$  0.26  &  1.92$\pm$  0.40  &  0.56$\pm$  0.10  &  0.26$\pm$  0.05  &  39.6$\pm$  18.8\\
     Enuc. 9   & 170.3$\pm$   8.8  & 301.2$\pm$  15.1  &  2.25$\pm$  0.25  &  3.31$\pm$  0.36  &  3.03$\pm$  0.61  &  0.95$\pm$  0.16  &  0.46$\pm$  0.08  &  58.8$\pm$  19.9
\enddata
\tablecomments{$^{\dagger}$ Corrected for stellar emission.  Observations were made using the following observatories and instruments: 
{\it Spitzer}-IRAC (8\,$\mu$m), {\it Spitzer}-MIPS (24\,$\mu$m), {\it Herschel}-PACS (70, 100, 160\,$\mu$m), {\it Herschel}-SPIRE (250, 350\,$\mu$m), and SCUBA (850\,$\mu$m).}
\end{deluxetable*}

\begin{deluxetable*}{ccccc}
\tablecaption{H$\alpha$ and UV Photometry \label{tbl-3}}
\tabletypesize{\scriptsize} 
\tablewidth{0pt}
\tablehead{
\colhead{ID}  & 
\colhead{$f_{\rm H\alpha}^{\dagger}$} & \colhead{$f_{\nu}$ (1528\,\AA)}  & \colhead{$f_{\nu}$ (2271\,\AA)}  & \colhead{$E(B-V)^{\ddagger}$}\\ 
\colhead{} &
\colhead{($10^{-12}$\,erg\,s$^{-1}$\,cm$^{-2}$)} & \colhead{(mJy)} & \colhead{(mJy)} & \colhead{(mag)}
}
\startdata
     Nucleus   &  1.09$\pm$  0.22  &  0.07$\pm$  0.01  &  0.42$\pm$  0.06  &0.342\\
     Enuc. 1   &  0.47$\pm$  0.09  &  1.25$\pm$  0.19  &  2.49$\pm$  0.37  &0.345\\
     Enuc. 2   &  2.01$\pm$  0.40  &  5.81$\pm$  0.87  &  9.58$\pm$  1.44  &0.346\\
     Enuc. 3   &  0.80$\pm$  0.16  &  2.52$\pm$  0.38  &  4.44$\pm$  0.67  &0.343\\
     Enuc. 4   &  0.61$\pm$  0.12  &  1.65$\pm$  0.25  &  3.30$\pm$  0.50  &0.343\\
     Enuc. 5   &  0.45$\pm$  0.09  &  1.04$\pm$  0.16  &  2.06$\pm$  0.31  &0.335\\
     Enuc. 6   &  1.09$\pm$  0.22  &  1.80$\pm$  0.27  &  3.75$\pm$  0.56  &0.344\\
     Enuc. 7   &  1.19$\pm$  0.24  &  1.93$\pm$  0.29  &  4.07$\pm$  0.61  &0.342\\
     Enuc. 8   &  0.36$\pm$  0.07  &  0.23$\pm$  0.04  &  0.67$\pm$  0.10  &0.342\\
     Enuc. 9   &  0.76$\pm$  0.15  &  1.16$\pm$  0.17  &  2.44$\pm$  0.37  &0.342
\enddata
\tablecomments{$^{\dagger}$ H$\alpha$ imaging obtained with the KPNO 0.9\,m by \citet{af98}.  
$^{\ddagger}$ Galactic extinction taken from \citep{ds98} used to correct the H$\alpha$ and GALEX FUV and NUV flux densities assuming $A_{V}/E(B-V)=3.1$ and the modeled extinction curves of \citet{wd01,bd03}.}  
\end{deluxetable*}

\section{Data and Photometry}
NGC\,6946 is a nearby \citep[$d \approx 6.8$~Mpc;][]{ik00} late-type spiral galaxy harboring a mild starburst nucleus \citep{th83,ball85}.  
Recently, \citet{ejm10} targeted 10 star-forming regions within NGC\,6946, including the nucleus, using the 100~m Robert C. Byrd Green Bank Telescope (GBT) in the Ka-band ($26-40$\,GHz).  
These regions were selected because they have existing and forthcoming mid- and far-infrared spectroscopic data collected as part of the {\it Spitzer} Infrared Nearby Galaxies Survey \citep[SINGS;][]{rck03} and the project, Key Insights on Nearby Galaxies: a Far-Infrared Survey with {\it Herschel} (KINGFISH; PI. R.C.~Kennicutt).  
The position of each source is given in Table \ref{tbl-1}.  
In Figure \ref{fig-1} we show the location of each region on the 24\,$\mu$m image of NGC\,6946; the corresponding circle diameters are 25\arcsec, which matches the FWHM of our lowest resolution data for the present multiwavelength SFR comparison.  
This projects to a physical scale of $\approx0.8$\,kpc at the distance of NGC\,6946.  


\subsection{Radio Data}  
Radio data at 1.4, 1.5, 1.7, 4.9, 8.5, and 33\,GHz used here are the same as the data presented in \citet{ejm10}, where a more detailed description of the observations and spectral decomposition of the thermal and non-thermal components can be found.  
The 1.4\,GHz radio map ($14\arcsec \times12\farcs5$ beam) comes from the Westerbork Synthesis Radio Telescope (WSRT)-SINGS survey \citep{rb07}.  
Imaging at 1.5, 1.7, 4.9, and 8.5\,GHz ($15\arcsec \times 15\arcsec$ beam) all come from \citet{beck07}.  
We note that the 4.9 and 8.5\,GHz radio data included single-dish measurements for short-spacing corrections.  

Observations at 33\,GHz were taken on 2009 March 21 using the Caltech Continuum Backend (CCB) on the GBT.  
The CCB simultaneously measures the entire Ka bandwidth over 4 equally spaced frequency channels.  
The beamswitched signal is synchronously read out and demodulated to remove atmospheric fluctuation and/or gain variations.  
Reference beams are measured by nodding 1\farcm3 away from the source; details about the reference beam locations can be found in Figure 1 of \citet{ejm10}.  
The average FWHM of the GBT beam in the Ka-band was $\approx$25\arcsec\ among our sets of observations.  
A detailed description on the performance of the CCB receiver, the data reduction pipeline, and error estimates are given in \citet{bm09}.  

\subsection{UV, Optical, and IR Data}
{\it GALEX} far-UV (FUV; 1528~\AA) and near-UV (NUV; 2271~\AA) data of NGC\,6946 were taken from the {\it GALEX} archive and will be included in the {\it GALEX} Large Galaxy Atlas (M. Seibert et al. 2011, in preparation).  
The calibration uncertainty for these data is $\approx$15\% in both bands.  
The H$\alpha$ image of NGC\,6946, obtained using the Tek $2048\times2048$ CCD on the KPNO 0.9~m ($23\arcmin \times 23\arcmin$ FOV and $0\farcs69$~pixel$^{-1}$), was taken from \citet{af98}, and has a calibration uncertainty of $\approx$20\%.  
The 8, 24, and 850\,$\mu$m data used here were included in the SINGS fifth data release,\footnote{SINGS data products can be found at
http://irsa.ipac.caltech.edu/data/SPITZER/SINGS/.} 
and have calibration uncertainties of 5, 5, and 15\%, respectively.  
Regions 2, 3, and 5 were not covered by the SCUBA 850\,$\micron$ maps, while regions 4 and 8 were near the edges of the maps.    
We note that the 8\,$\micron$ map was corrected for stellar light using a scaled 3.6\,$\micron$ image following \citet{gxh04}.  
We do not use the {\it Spitzer} 70\,$\mu$m or the SCUBA 450\,$\mu$m data, also included in the SINGS fifth data release, in lieu of new {\it Herschel} data taken as part of KINGFISH.  

We make use of the {\it Herschel} \citep{glp10} data obtained by the KINGFISH project having resolutions better than the 25\arcsec\ beam of our GBT data which include PACS 70, 100, and 160\,$\mu$m imaging, along with SPIRE 250 and 350\,$\mu$m imaging.  
Both PACS and SPIRE imaging were reduced using the {\it Herschel} Interactive Processing Environment \citep[HIPE;][]{hipe10} software package, version 3.0.1510.  
The current calibration has uncertainties of $\sim$10, 10, and 20\% for the 70, 100 and 160\,$\mu$m bands, respectively \citep{ap10}. 
The uncertainty in the SPIRE 250 and 350\,$\mu$m data is taken to be 15\%, which is the quadrature sum of the 15\% absolute calibration uncertainty and a 1\% uncertainty in the size of the beam;  
following the current recommendation by the SPIRE Observer's Manual, we assume beam sizes at 250 and 350\,$\mu$m of 426 and 771 arcsec$^{2}$, respectively.  
Calibration correction factors of 1.02 and 1.05 were applied to the 250, 350\,$\mu$m data as suggested in the SPIRE Scan-Map AOT Observer's manual.

\subsection{Photometry}
The photometry was carried out on all UV, optical, infrared, and radio maps after first cropping each image to a common field-of-view and regridding to a common pixel scale.  
To accurately match the photometry of our images to the GBT measurements, maps were convolved to the resolution of the GBT beam in the Ka-band following the image registration description given in \citet{ejm10}.  
Registration of the {\it Herschel} data was carried out in a similar way as the registration of the {\it Spitzer} data.  
Each of the {\it Herschel} images were first convolved to the resolution of the 350\,$\mu$m data using custom smoothing kernels as described in \citet{kdg08} and updated
by those authors for {\it Herschel}, CLEANed to suppress side-lobe structure in the PSF, and then restored using a Gaussian beam having a FWHM of 25\arcsec.  
Flux densities at each wavelength were measured by taking the surface brightness at the location of each GBT pointing and multiplying by the effective area of the beam.  
In the case of the UV and H$\alpha$ photometry, we correct each region for Milky Way extinction using \citet{ds98} assuming $A_{V}/E(B-V)=3.1$ and the modeled extinction curves of \citet{wd01,bd03}.  

Unlike \citet{ejm10}, we subtract a local background estimate from our photometry within the vicinity of each star-forming region to remove any diffuse emission component that is most likely unassociated with ongoing star formation.  
Results using the photometry without subtracting local backgrounds are discussed in $\S$\ref{sec-glob} and in the Appendix.  
The local background estimates were measured by placing four 25\arcsec\ diameter apertures at a distance of 1.5 times the beam FWHM (i.e., 37\farcs5) away from the center of the source positions in each of the four Cardinal directions.  
The median surface brightness among the pixels within these four apertures (typically 100 pixels in total) was then multiplied by the effective area of the beam to get an estimate of the local diffuse background emission.  
While such a local background subtraction for our 33\,GHz GBT single-pointing observations is not possible, we note that those observations were taken using 1\farcm3 nods, thus reference beams were typically measured on diffuse parts of the galaxy and likely provide a reasonable measure of the local background.  
Furthermore, at 33\,GHz, we do not expect there to be a significant underlying diffuse component if the emission is in fact dominated by free-free emission.     

Photometric uncertainties were conservatively estimated by taking the quadrature sum of the calibration and background uncertainties, along with the RMS noise for each image.  
The results from our radio, infrared, and UV/H$\alpha$ photometry, along with 1-$\sigma$ uncertainties, are given in Tables \ref{tbl-1}, \ref{tbl-2}, and \ref{tbl-3}, respectively.  
The local background values, along with the average fractional contribution of the background emission at each waveband and region, are given in the Appendix.    
For the cases in which an over subtraction of signal in the 33\,GHz observations has been estimated to be at the $\ga$15\% level due to reference beams landing on bright regions in the galaxy disk, we apply correction (`boosting') factors to the 33\,GHz flux densities \citep[see][]{ejm10}.  
These scale factors were derived based on the 8.5\,GHz and 24\,$\mu$m images, and are approximately 2.55, 1.17, and 1.25 for Enuc. 1, 5, and 8, respectively.  
We note that these values are $\la2$\% smaller than those reported by \citet{ejm10} due to including local background subtractions.  

\subsection{Modeling the Infrared-to-Radio Spectra}
\label{sec-modspec}
We fit the radio and infrared data independently given that the physical processes producing these emissions are distinct.  
The infrared (i.e., $8-850\,\mu$m) photometry are fit by the spectral energy distribution (SED) models of \citet{dh02}; associated total infrared (IR: $8-1000\,\mu$m) luminosities from integrating the best-fit SEDs for each region are given in Table \ref{tbl-4}.  
Fitting errors in the IR luminosity determinations were estimated by a standard Monte Carlo approach using the photometric uncertainties of the input flux densities.  

The radio data are fit by varying a combination of thermal (free-free) and non-thermal (synchrotron) emission components which scale as $S_{\nu}^{\rm T}  \propto \nu^{-\alpha_{\rm T}}$ and $S_{\nu}^{\rm NT}  \propto \nu^{-\alpha_{\rm NT}}$, respectively, where $\alpha_{\rm T}= 0.1$ and $\alpha_{\rm NT}$ (see Table \ref{tbl-4} for estimated values) are the thermal and non-thermal spectral indices, respectively.   
The only exception to this 
is for extranuclear region 4 where the thermal and non-thermal emission were fit only to the radio data at $\nu < 10$\,GHz and an additional component \citep[i.e., spinning dust emission; ][]{ahd09} was used to fit the excess 33\,GHz emission \citep[see][for details]{ejm10}.  
Recent follow-up observations of this region at 16.5 GHz using the Arcminute Microkelvin Imager have confirmed a rising spectrum between 8.5 and 15 GHz \citep{as10}.
The excess emission accounts for $\approx$56\% of the total 33\,GHz flux density from extranuclear region 4 and has been subtracted out for the present analysis.  
We note that the quantitative results presented here are significantly different from what is given in \citet{ejm10} as those authors did not subtract out a local background estimate and included a spinning dust component for star-forming regions in addition to extranuclear region 4.  
Results associated with fitting the photometry without subtracting a local background can be found in $\S$\ref{sec-glob} and the Appendix.  

Before varying the free-free and non-thermal components to fit the observed radio spectra, we first estimate the non-thermal spectral index for each source.  
The non-thermal emission is proportional to the ratio of energy losses of CR electrons from synchrotron emission to total energy losses.  
In our model, we account for energy losses to CR electrons arising from synchrotron radiation, inverse Compton (IC) scattering, ionization, bremsstrahlung, and escape through an empirical prescription.  
Additional losses through advection from a galactic scale wind are ignored.  
A much more detailed description on the construction of the modeled radio spectra can be found in \citet{ejm09c}.  

Synchrotron losses are proportional to $U_{B} = B^{2}/(8\pi)$, the magnetic field energy density, while IC losses are proportional to $U_{\rm rad}$, the radiation field energy density.  
We use magnetic field strengths estimated from the 1.4\,GHz radio continuum surface brightnesses following the revised minimum energy calculation of \citet[][see Table \ref{tbl-4}]{bk05}.  
For this calculation we assume a proton-to-electron number density ratio of $K_{0} \approx 100\pm 50$ and a pathlength of $l \approx 1\pm 0.5$\,kpc.  
The estimated $B_{\rm min}$ values range between $\approx10-16\,\mu$G among the extranuclear star-forming regions and it is $\approx 30\,\mu$G for the nucleus.  
Assuming that the bolometric surface brightness of these dusty star-forming regions is well approximated by the IR surface brightness, we estimate $U_{\rm rad}\approx (2\pi/c) I_{\rm IR}$.  
These values range between $\approx 0.4 - 4.2\times10^{-12}$~ergs~cm$^{-3}$ among the extranuclear star-forming regions, and it is $\approx 41\times10^{-12}$ ~ergs~cm$^{-3}$ in the nucleus.  
We assume that the density of the interstellar medium (ISM) is $n_{\rm ISM} = 0.1$ and 10~cm$^{-3}$ for the extranuclear regions and nucleus, respectively.  
Taking these values, we make a rough estimate for the shape (i.e., spectral index) of the non-thermal radio spectrum.   
Once the non-thermal spectral index is fixed, the thermal and non-thermal components are varied to fit the radio observations.      

\begin{figure}
\scalebox{1.1}{
\plotone{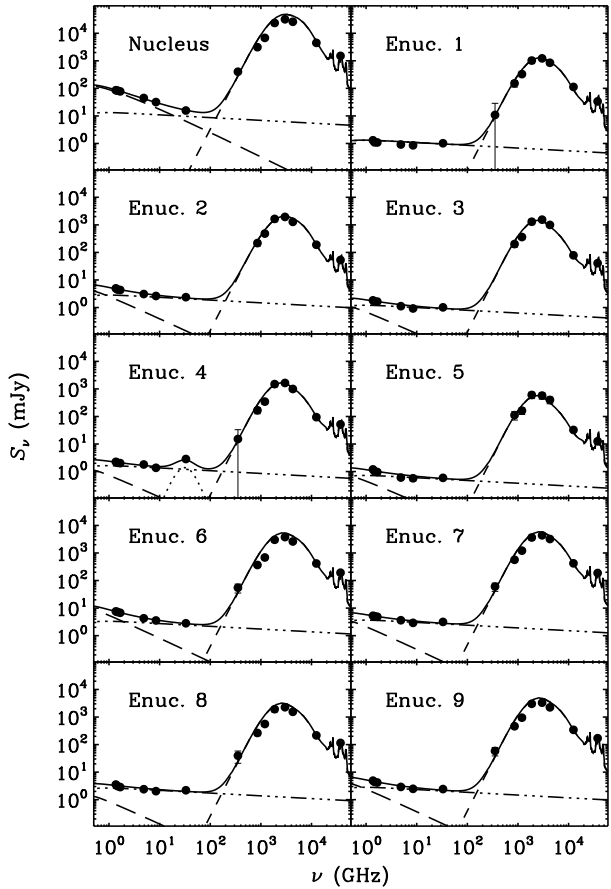}}
\caption{
The radio to infrared spectra of each star-forming region.  
The radio and infrared photometry are plotted as filled circles; error bars are shown, but are usually smaller than the size of the plotting symbol.  
Fits to the data using a combination of modeled radio spectra \citep{ejm09c} and infrared SED libraries of \citet{dh02} are shown as solid lines.  
The infrared dust emission is shown by a short-dashed line.  
The radio spectrum is modeled by 2 components, non-thermal synchrotron emission (long-dashed lines) and thermal (free-free) emission (triple-dot dashed line) except for the case of extranuclear region 4, where an additional component \citep[i.e., spinning dust emission;][dotted line]{ahd09} was needed to fit the excess emission at 33\,GHz \citep[see][for details]{ejm10}.  
\label{fig-2}}
\end{figure}

The radio to infrared spectra of each star-forming region are plotted in Figure \ref{fig-2}, along with the components of the fits. 
In Table \ref{tbl-4} we list the observed ($\alpha^{\rm obs}$) and (estimated) non-thermal radio spectral indices measured between 1.4 and 8.5\,GHz, along with the associated thermal radio fractions at 33\,GHz from our radio spectral fitting for each star-forming region.     
The observed indices are quite flat, being $\approx$0.2, on average, consistent with young H{\sc ii} regions, while the average non-thermal spectral index is $\approx$0.8.  
The average 33\,GHz thermal fraction is found to be $\approx$87\% among these star-forming regions.   

\section{SFR Calibrations}\label{sec-sfrcal}
To ensure that each SFR diagnostic can be compared fairly, 
we derive corresponding calibrations using Starburst99 \citep[][]{cl99} for a common IMF.  
We choose a Kroupa \citep{pk01} IMF, having a slope of $-1.3$ for stellar masses between $0.1-0.5~M_{\sun}$ and $-2.3$ for stellar masses ranging between $0.5-100~M_{\sun}$.  
These calibrations update those presented in \citet{rck98}.  
Only SFR diagnostics that are considered to be reliable independent of extinction are discussed.  

\subsection{UV, Optical, and IR SFRs}
Assuming a solar metallicity and continuous star formation (i.e., a fixed SFR), Starburst99 stellar population models yield the following relation between the SFR and production rate of ionizing photons, $Q(H^{0})$, at an age of $\sim$100~Myr:  
\begin{equation}
\label{eq-Nly}
\left(\frac{\rm SFR}{M_{\sun}~{\rm yr^{-1}}}\right) = 7.29\times10^{-54}\left[\frac{Q(H^{0})}{s^{-1}}\right].  
\end{equation}
For reference, the coefficient relating the SFR and ionizing photon rate given in \citet{rck98} is a factor of $\approx$1.5 times larger than that above \citep[see][for a discussion on the origin of this difference]{dc07}.  
We choose a constant SFR over a timescale of $\sim$100~Myr for all estimators as we are sampling star-forming complexes on the scales of $\approx$0.8\,kpc that likely include a range of ages among individual star formation sites.     
Consequently, below certain spatial scales (i.e., $\la$ a few hundred \,pc; the size of individual giant H{\sc ii} regions) these calibrations will likely break down as the assumption of spatial averaging is no longer valid.  

The ionizing photon rate can of course be expressed as a (extinction corrected) H~recombination line flux, such that for Case B recombination, and assuming an electron temperature $T_{\rm e} = 10^{4}$~K, the H$\alpha$ recombination line strength is related to the SFR by 
\begin{equation}
\label{eq-sfrha}
\left(\frac{\rm SFR_{H\alpha}}{M_{\sun}~{\rm yr^{-1}}}\right) = 5.37\times10^{-42}\left(\frac{L_{{\rm H}\alpha}}{\rm erg~s^{-1}}\right).  
\end{equation}
The above equation indicates that the SFR is directly proportional to the H$\alpha$ line luminosity, assuming that a constant fraction of the ionized H~atoms will emit an H$\alpha$ photon as they recombine, and that the extinction correction is accurate.
However, if a significant fraction of ionizing photons are absorbed by dust, the above equation will underestimate the SFR, and the $=$ sign should be replaced by $\geq$.  
Since the ionizing flux comes from very massive stars with lifetimes $\la$10~Myr, we note that the coefficients in Equations \ref{eq-Nly} and \ref{eq-sfrha} are nearly independent of starburst age under the assumption of continuous star formation so long at it is $\ga$10~Myr.  
Accordingly, such measurements sample the current (i.e., $\sim$10~Myr) star formation activity.  

The integrated UV spectrum is dominated by young stars, making it a sensitive probe of recent \citep[$\sim$$10-100$~Myr;][]{rck98,dc05,ss07} star formation activity.  
We convolve the output Starburst99 spectrum with the {\it GALEX} FUV transmission curve to obtain the following conversion between SFR and FUV luminosity, 
\begin{equation}
\label{eq-sfrfuv}
\left(\frac{\rm SFR_{FUV}}{M_{\sun}~{\rm yr^{-1}}}\right) = 4.42\times10^{-44}\left(\frac{L_{\rm FUV}}{\rm erg~s^{-1}}\right).
\end{equation}

To derive a calibration for the total infrared (IR; $8-1000\,\mu$m), we make the assumption that the entire Balmer continuum is absorbed and re-radiated by dust and that the dust emission is optically thin.  
Integrating the output Starburst99 spectrum over this wavelength range (i.e., $912 < \lambda < 3646$~\AA) results in the following (predicted) relation between the IR emission and current SFR:  
\begin{equation}
\label{eq-sfrir}
\left(\frac{\rm SFR_{IR}}{M_{\sun}~{\rm yr^{-1}}}\right) = 3.88\times10^{-44}\left(\frac{L_{\rm IR}}{\rm erg~s^{-1}}\right).  
\end{equation}


Since a non-negligible fraction of the far-infrared continuum may be heated by an older stellar population \citep[e.g.,][]{st92,wg96,gjb10}, it is likely more physical to only relate warmer dust emission to the current SFR \citep[e.g.,][]{gxh04}.  
Thus, we also include a monochromatic 24\,$\mu$m based SFR diagnostic, for which there are a number of calibrations in the literature \citep[e.g.,][]{hwu05, ppg06,aah06,dc07,mr07,ynz08,gr09}.  
A detailed comparison among each of these relations can be found in \citet{dc10}.  
For simplicity, we only choose a single calibration to use in our comparison \citep[i.e.,][calibrated for a Kroupa IMF]{mr07}, such that
\begin{equation}
\label{eq-sfr24}
\left(\frac{\rm SFR_{24\micron}}{M_{\sun}~{\rm yr^{-1}}}\right) = 5.58\times10^{-36}\left[\frac{\nu L_{\nu} (24~\micron)}{\rm erg~s^{-1}}\right]^{0.826}, 
\end{equation}
as this relation has been calibrated over the largest range in 24\,$\mu$m luminosity (i.e. $1\times10^{38}~{\rm erg~s^{-1}} \la \nu L_{\nu} (24~\micron) \la 3 \times 10^{44}~{\rm erg~s^{-1}}$).  

Given that not all of the UV/optical photons will be absorbed and re-radiated by dust, a series of new empirical calibrations based on the linear combination of observed 24\,$\mu$m (obscured star formation) and H$\alpha$ (unobscured star formation) luminosities have been developed \citep[e.g.,][]{dc07,rck07,rck09,ynz08}.  
Based on the analysis of \citet{dc07}, we adopt the following relation for our sample of star-forming regions to obtain extinction corrected H$\alpha$ line luminosities: 
\begin{equation}
\label{eq-hacorr}
\left(\frac{L_{{\rm H}\alpha}^{\rm corr}}{\rm erg~s^{-1}}\right) = \left[\frac{L_{{\rm H}\alpha}^{\rm obs} + 0.031 \nu L_{\nu}(24~\micron)}{\rm erg~s^{-1}}\right].  
\end{equation}
While this relation has been calibrated for H{\sc ii} regions, we note that it has also been found to be appropriate for entire galaxies having \(4\times10^{42} \leq \nu L_{\nu}(24~\micron) < 5\times10^{43}~{\rm erg~s^{-1}}\).  
For galaxies having \(\nu L_{\nu}(24~\micron) < 4\times10^{42}~{\rm erg~s^{-1}}\), this relation is optimized by replacing the coefficient 0.031 by 0.020 \citep{dc10}, which is the calibration derived for normal star-forming galaxies by \citet{rck09}. 
Combining Equation \ref{eq-sfrha} with Equation \ref{eq-hacorr} results in the following SFR relation, 
\begin{equation}
\label{eq-sfrmix}
\left(\frac{\rm SFR_{mix}}{M_{\sun}~{\rm yr^{-1}}}\right) = 5.37\times10^{-42}\left[\frac{L_{{\rm H}\alpha}^{\rm obs} + 0.031 \nu L_{\nu}(24~\micron)}{\rm erg~s^{-1}}\right].  
\end{equation}

Similarly, by combining the SFR estimates from the total IR and (observed) UV emission, one can account for the obscured and unobscured emission contributing to the total (bolometric) SFR, which we define as: 
\begin{equation}
\label{eq-sfrtot}
{\rm SFR_{tot}}  =  {\rm SFR_{FUV}} + {\rm SFR_{IR}}. 
\end{equation}
Using the above calibrations, this relation can be expressed as a linear combination of the UV and total IR emission such that, 
\begin{equation}
\label{eq-sfrtot2}
\left(\frac{\rm SFR_{tot}}{M_{\sun}~{\rm yr^{-1}}}\right) = 4.42\times10^{-44} \left[\frac{L_{\rm FUV} + 0.88 L_{\rm IR}} {\rm erg~s^{-1}} \right].  
\end{equation}
This diagnostic is often used to characterize SFRs from galaxies in both low- \citep[e.g.,][]{jip06,vb07,vb11} and high-$z$ \citep[e.g.,][]{jip04,de07,ed07a,nr10} studies.  
However, as already mentioned above, the presented calibrations will depend on each complex including a range of ages among individual star formation sites. 
Therefore, calibrations such as these will have increasing uncertainties when applied to a single star-forming region depending on how well the IR and UV are measuring emission from recent star formation, not to mention the fact that the IR and UV are sensitive to star formation on different timescales (i.e., $\sim$10~Myr versus $\sim$$10-100$~Myr, respectively).  
It is also worth noting that the derived coefficient for scaling the IR luminosity given above is larger than that from empirical studies of starburst galaxies which report a coefficient closer to $\sim$0.6  \citep[e.g.,][]{gm99, dc01}.  

\subsection{Radio SFRs}
\label{sec-sfrrad}
Radio continuum emission from galaxies is generally composed of two optically-thin components: 
non-thermal synchrotron emission associated with CR electrons accelerated in a galaxy's magnetic field, and thermal bremsstrahlung (free-free) emission around massive star-forming regions.  
The origin of each of these components lies in the process of massive star formation.  

At high radio frequencies, where $\tau \ll 1$, the ionizing photon rate is directly proportional to the thermal spectral luminosity, $L_{\nu}^{\rm T}$, varying only weakly with electron temperature $T_{\rm e}$ \citep{rr68}, such that 
\begin{equation}
\label{eq-Nlyrad}
\begin{split}
\left[\frac{Q(H^{0})}{s^{-1}}\right] &= 6.3\times10^{25} \\ 
&\left(\frac{T_{\rm e}}{10^{4}~{\rm K}}\right)^{-0.45} \left(\frac{\nu}{\rm GHz}\right)^{0.1} \left(\frac{L_{\nu}^{\rm T}}{\rm erg~s^{-1}~Hz^{-1}}\right).  
\end{split}
\end{equation}
As with the H~recombination line fluxes, it is again worth noting that $Q(H^{0})$, and consequently the SFR, may in fact be underestimated by the free-free emission if  a significant fraction of ionizing photons are absorbed by dust; in this case the $=$ sign in the above equation should be replaced by $\geq$.   
By combining Equations \ref{eq-Nly} and \ref{eq-Nlyrad}, one can derive a relation between the SFR and  thermal radio emission: 
\begin{equation}
\label{eq-sfrt}
\begin{split}
\left(\frac{\rm SFR_{\nu}^{T}}{M_{\sun}~{\rm yr^{-1}}}\right) &= 4.6\times10^{-28}\\
&\left(\frac{T_{\rm e}}{10^{4}~{\rm K}}\right)^{-0.45} \left(\frac{\nu}{\rm GHz}\right)^{0.1} \left(\frac{L_{\nu}^{\rm T}}{\rm erg~s^{-1}~Hz^{-1}}\right).  
\end{split}
\end{equation}
Using a Kroupa IMF that extends down to 0.1\,$M_{\sun}$ results in a coefficient that is a factor of $\sim$2.5 times larger than that found in the relation between the SFR and free-free radio emission given in \citet{jc92}, and $\sim$25\% smaller than that given in \citet{hrs06}.    

At lower radio frequencies, which are typically dominated by non-thermal synchrotron emission, calibrations between the supernova rate, and thus the SFR, have been developed.  
From the output of Starburst99, which assumed a supernova cut-off mass of 8\,$M_{\sun}$, we find that the total core-collapse supernova rate, $\dot{N}_{\rm SN}$, is related to the star formation rate by,   
\begin{equation}
\label{eq-sfrqsnr}
\left(\frac{\rm SFR}{M_{\sun}~{\rm yr^{-1}}}\right) =  86.3 \left(\frac{\dot{N}_{\rm SN}}{\rm~yr^{-1}}\right).
\end{equation}
Work comparing the non-thermal spectral luminosity with the supernova rate in the Galaxy has yielded an empirical calibration such that
\begin{equation}
\label{eq-lntqsnr}
\left(\frac{L_{\nu}^{\rm NT}}{\rm erg~s^{-1}~Hz^{-1}}\right) = 1.3\times10^{30} \left(\frac{\dot{N}_{\rm SN}}{\rm yr^{-1}}\right) \left(\frac{\nu}{\rm GHz}\right)^{-\alpha^{\rm NT}}
\end{equation}
\citep{gtam82,cy90}.
By combining Equations \ref{eq-sfrqsnr} and \ref{eq-lntqsnr}, we can express the star formation rate as a function of the non-thermal radio emission where 
\begin{equation}
\label{eq-sfrnt}
\left(\frac{\rm SFR_{\nu}^{NT}}{M_{\sun}~{\rm yr^{-1}}}\right) =6.64\times10^{-29}  \left(\frac{\nu}{\rm GHz}\right)^{\alpha^{\rm NT}}\left(\frac{L_{\nu}^{\rm NT}}{\rm erg~s^{-1}~Hz^{-1}}\right).
\end{equation}
We note that using the Kroupa IMF that extends down to 0.1\,$M_{\sun}$ results in a coefficient that is a factor of $\sim$3.5 times larger than that used in the expression relating the SFR and non-thermal radio emission by  \citet{jc92}, and $\sim$50\% smaller than that given in \citet{hrs06}.    

\begin{deluxetable*}{ccccccc}
\tablecaption{Derived Parameters \label{tbl-4}}
\tabletypesize{\scriptsize} 
\tablewidth{0pt}
\tablehead{
\colhead{ID}  & 
\colhead{$f_{\rm T}^{\rm 33\,GHz}$} & 
\colhead{$\alpha^{\rm obs \dagger}$} & \colhead{$\alpha^{\rm NT \dagger}$} &
\colhead{$L_{\rm IR}$}& \colhead{$B_{\rm min}^{\ddagger}$} & \colhead{$T_{\rm e}$}\\  
\colhead{} &
\colhead{(\%)} & \colhead{} &\colhead{} & \colhead{($10^{8} L_{\sun}$)}&
\colhead{($\mu$G)} & \colhead{($10^{4}$\,K)}
}
\startdata
     Nucleus   & 62$\pm$ 6  &0.58$\pm$0.04  &0.74$\pm$0.05  &47.56$\pm$ 2.14& 30.3$\pm$  9.3& 0.42$\pm$ 0.08\\
     Enuc. 1   & 93$\pm$ 4  &0.10$\pm$0.01  &0.78$\pm$0.06  & 1.08$\pm$ 0.08& 10.1$\pm$  3.0& 1.36$\pm$ 0.31\\
     Enuc. 2   & 87$\pm$ 5  &0.29$\pm$0.02  &0.82$\pm$0.06  & 1.92$\pm$ 0.11& 14.3$\pm$  4.1& 0.77$\pm$ 0.27\\
     Enuc. 3   & 87$\pm$ 7  &0.23$\pm$0.02  &0.79$\pm$0.06  & 1.10$\pm$ 0.08& 11.0$\pm$  3.2& 0.86$\pm$ 0.31\\
     Enuc. 4   & 42$^{*}\pm$ 4  &0.20$\pm$0.01  &0.80$\pm$0.06  & 1.32$\pm$ 0.08& 11.6$\pm$  3.4& 1.78$\pm$ 0.59\\
     Enuc. 5   & 89$\pm$ 8  &0.24$\pm$0.02  &0.77$\pm$0.05  & 0.41$\pm$ 0.05&  9.9$\pm$  2.9& 1.07$\pm$ 0.44\\
     Enuc. 6   & 86$\pm$ 5  &0.38$\pm$0.03  &0.83$\pm$0.06  & 4.72$\pm$ 0.25& 16.0$\pm$  4.6& 1.04$\pm$ 0.23\\
     Enuc. 7   & 85$\pm$ 5  &0.23$\pm$0.02  &0.83$\pm$0.06  & 4.86$\pm$ 0.26& 14.5$\pm$  4.2& 1.14$\pm$ 0.25\\
     Enuc. 8   & 88$\pm$ 5  &0.18$\pm$0.01  &0.81$\pm$0.06  & 2.66$\pm$ 0.14& 13.1$\pm$  3.8& 2.80$\pm$ 0.54\\
     Enuc. 9   & 85$\pm$ 5  &0.28$\pm$0.02  &0.83$\pm$0.06  & 4.12$\pm$ 0.21& 14.3$\pm$  4.1& 1.22$\pm$ 0.26
\enddata
\tablecomments{$^{\dagger}$ The observed and non-thermal radio
  spectral indices measured between 1.4 and 8.5\,GHz.   
$^{\ddagger}$ Calculated using the total 1.4\,GHz flux
  densities and the revised minimum energy calculation of \citet{bk05}.    
$^{*}$ Calculated using the observed 33\,GHz flux density before subtracting out the $\approx$56\% contribution thought to arise from anomalous dust emission (see $\S$\ref{sec-modspec}). }
\end{deluxetable*}

Since the observed radio continuum emission is comprised of both free-free and synchrotron emission, we can combine Equations \ref{eq-sfrt} and \ref{eq-sfrnt} to construct a single expression for the SFR from the total radio continuum emission at a given frequency such that: 
\begin{equation}
\label{eq-sfrrad}
\begin{split}
\left(\frac{\rm SFR_{\nu}}{M_{\sun}~{\rm yr^{-1}}}\right) &= 10^{-27}  
\left[2.18 \left(\frac{T_{\rm e}}{10^{4}~{\rm K}}\right)^{0.45} \left(\frac{\nu}{\rm GHz}\right)^{-0.1}\right. + \\
&\left.15.1 \left(\frac{\nu}{\rm GHz}\right)^{-\alpha^{\rm NT}}\right]^{-1} \left(\frac{L_{\nu}}{\rm erg~s^{-1}~Hz^{-1}}\right).  
\end{split}
\end{equation}
This equation essentially weights the observed radio continuum luminosity based on the expected thermal fraction at a given frequency.  

The most widely used radio SFR calibration, however, is the result of the tight, empirical FIR-radio correlation \citep{de85,gxh85}:   
\begin{equation}
\label{eq-q}
q_{\rm IR} \equiv \log~\left(\frac{L_{\rm IR}}{3.75\times10^{12}L_{\rm 1.4GHz}}\right).  
\end{equation}
While classically expressed using the integrated ({\it IRAS}-based) FIR luminosity, we use the total IR luminosity resulting in an average value of $q_{\rm IR} = 2.64\pm 0.26$~dex for a sample of star-forming galaxies in the local universe \citep{efb03}; this value is 0.30~dex larger than the average value of $2.34\pm0.26$~dex found when using the FIR luminosity \citep{yrc01}.  
Thus, one can express the 1.4\,GHz luminosity as a SFR by combining Equation \ref{eq-q} with Equation \ref{eq-sfrir}: 
\begin{equation}
\label{eq-sfrq}
\left(\frac{\rm SFR_{\rm 1.4GHz}}{M_{\sun}~{\rm yr^{-1}}}\right) = 6.35\times10^{-29}\left(\frac{L_{\rm 1.4GHz}}{\rm erg~s^{-1}~Hz^{-1}}\right).   
\end{equation}
It is worth stating here that this calibration is based on the FIR-radio correlation, which was established for globally integrated FIR and radio continuum properties of galaxies, and thus applies to galaxies as a whole.  
Most of the non-thermal radio emission from galaxies arises from electrons that have diffused away from their acceleration sites in supernova remnants into quiescent regions void of current star formation.  
In fact, within individual H{\sc ii} regions, there may be little non-thermal emission present \citep[e.g.,][]{cy90}.  
Thus, locally near star-forming complexes, we expect the non-thermal emission per unit star formation to be low given such a calibration as suggested by studies of the local FIR-radio correlation \citep[e.g.,][]{ejm06a,ah06,fat07a}

\begin{figure}
\begin{center}
\scalebox{1.2}{
\plotone{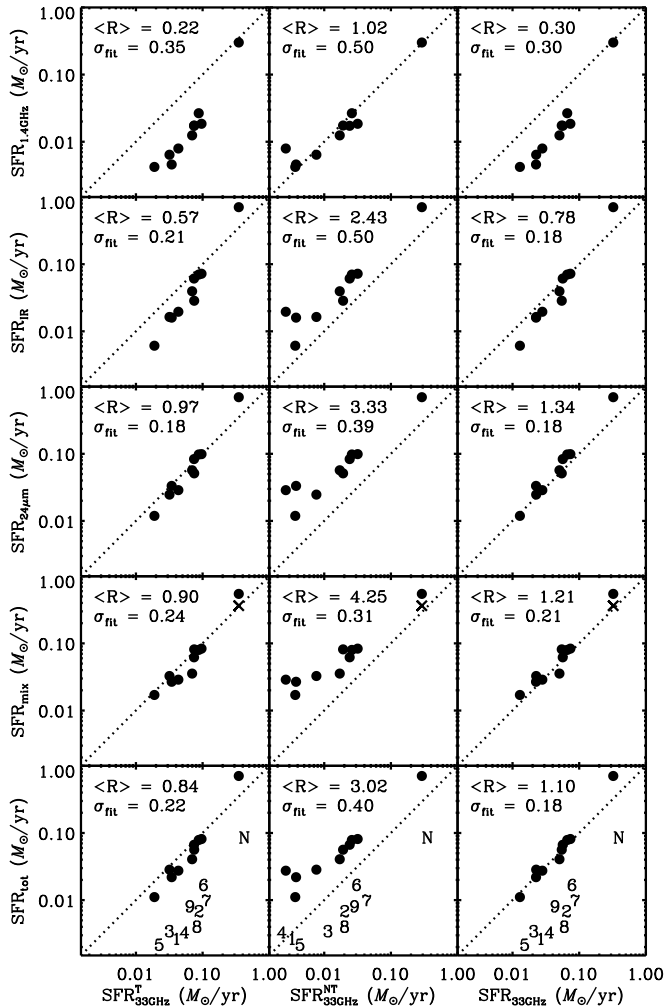}}
\caption{
A comparison of the various extinction independent SFR diagnostics described in $\S$\ref{sec-sfrcal} (i.e., 1.4\,GHz via the FIR-radio correlation, total IR, 24\,$\mu$m, H$\alpha+24\,\mu$m -- mix, and UV + IR -- tot), plotted from left to right against the 33\,GHz free-free, non-thermal, and total radio SFRs, respectively.  
To help guide the eye, the one-to-one line is shown as a dotted line in each panel.  
The points are also  identified per column in the bottom row of panels by the extranuclear region number and an `N' for the nucleus.  
In the upper left corner of each panel, the median ratio of the ordinate to abscissa ($<$R$>$), along with the residual dispersion about an ordinary least squares fit to each trend ($\sigma_{\rm fit}$), are given.  
In the plots comparing the H$\alpha+24\,\mu$m (mix) and 33\,GHz SFRs, we additionally plot the nucleus using the H$\alpha+24\,\mu$m calibration for entire galaxies \citep{rck09} as a cross.  
\label{fig-3}}
\end{center}
\end{figure}

\section{Comparison of SFR Diagnostics}
In the following section we compare SFRs estimated via the various diagnostics described in $\S$\ref{sec-sfrcal}.  
As previously mentioned, we only include SFR diagnostics whose reliability is considered to be fairly insensitive to extinction.  
SFRs estimated by free-free emission at 33\,GHz are assumed to be the most accurate, and thus act as the reference SFR for the comparison.  
This assumption is made as the thermal fraction at 33\,GHz is found to be $\approx$87\%, on average.  
The various SFR diagnostics are plotted in Figure \ref{fig-3} from left to right against the 33\,GHz free-free, non-thermal, and total radio SFRs, respectively.  
The median ratio of the ordinate to abscissa ($<$R$>$), along with the residual dispersion after subtracting out an ordinary least squares fit to the data ($\sigma_{\rm fit}$), are given in the top left corners of each panel.  

\subsection{Non-thermal Radio Emission: 1.4\,GHz Emission and the FIR-Radio Correlation}
In the top 3 panels of Figure \ref{fig-3}, SFRs derived using the observed 1.4\,GHz flux densities (along with the FIR-radio correlation) from each star-forming region are plotted against the three 33\,GHz derived estimates.  
Except for the starbursting nucleus, the 1.4\,GHz SFRs appear to significantly underestimate (i.e., by a factor of $\sim$5) the actual SFR as measured by the free-free emission at 33\,GHz.  
As stated at the end of $\S$\ref{sec-sfrrad}, this discrepant behavior is expected given that the FIR-radio correlation applies to galaxies as a whole.  
The agreement for the nucleus likely arises from the fact that the 33\,GHz thermal fraction (i.e., $\sim$62\%) is significantly smaller than that of the extranuclear star-forming regions, being much more similar to what is found for globally integrated measurements of galaxies   \citep[e.g.,][]{jc92}.     
Differences between the radio and optical properties (e.g., 1.4\,GHz to H$\alpha$ ratios) of galaxy nuclei and H{\sc ii} regions has been noted in the past by a number of authors \citep[e.g.,][]{ik80,heck83,vdh88}.  
These differences appear to be due to a larger non-thermal fraction in the nucleus relative to what is estimated for the extranuclear regions as previously suggested  by \citet{rck89}, rather than simply higher optical extinctions.  

Comparing the 1.4\,GHz SFRs with the 33\,GHz SFRs calculated using the total 33\,GHz flux densities (and Equation \ref{eq-sfrrad}), we again find that the 1.4\,GHz SFR calibrations are underestimating the SFR as measured by the total 33\,GHz flux density among the extranuclear regions.  
This is similar to the discrepancy between the 1.4\,GHz and 33\,GHz free-free SFRs, and most likely due to the fact that the thermal fractions are large at 33\,GHz among these star-forming regions.  
On the other hand, the 1.4\,GHz SFRs are nearly equal to those measured using the non-thermal 33\,GHz flux densities among both the extranuclear regions and the nucleus.  
This is not too surprising given that the 1.4\,GHz flux density is dominated by non-thermal emission, and suggests that our thermal decomposition is in fact fairly robust.  
The only region where there is a significant difference is extranuclear region 4, however, the thermal/non-thermal decomposition here is complicated by the presence of an anomalous emission component.    

\subsection{Total Infrared}
The second row of panels in Figure \ref{fig-3} compares the IR-based SFRs to those at 33\,GHz.  
We find that the IR-based SFRs typically underestimate the SFRs derived from the free-free emission among the extranuclear regions by an average factor of $\sim$2.  
There may also be a slight trend of increased discrepancy between the two SFR indicators with decreasing infrared surface brightness.  
This discrepancy among the extranuclear regions likely arises from the subtraction of a local background together with the application of a SFR calibration that assumes all UV photons are absorbed by dust at the source, which may not be true, since a significant fraction of non-ionizing photons can escape the star-forming complexes, powering the diffuse interstellar dust emission.  
That is, the lower surface brightness extranuclear star-forming regions may become increasingly porous to UV photons.  
Thus, by subtracting a local background, which makes up a significant fraction of the cool/cold dust emission, the IR-based SFR calibration yields values that are significantly lower than the free-free emission estimates.  
Similar results are seen when comparing the IR and radio SFRs using the total 33\,GHz flux densities.  

On the other hand, the nucleus 
is a factor of $\sim$2 times larger than the estimate using the free-free or total 33\,GHz emission.  
One reason for the large discrepancy in the nuclear SFR could arise from a significant amount of dust being heated by stars that are not contributing significantly to the ionization of H atoms.  
Alternatively, this difference may arise from the capture of ionizing photons by dust, which would decrease the free-free emission at 33\,GHz and increase the IR emission.  
A detailed discussion about such scenarios is given in $\S$\ref{sec-nucdiff}.  

The IR-based SFRs are generally larger (i.e., by a factor of more than $\sim$2, on average) than the non-thermal 33\,GHz SFRs for all star-forming regions.  
This is consistent with the above 1.4\,GHz SFR comparison among the extranuclear regions which suggest that the SFR calibration based on the non-thermal radio emission typically underestimates the true SFR in such regions.    
This finding is also in agreement with studies of the resolved FIR-radio correlation within galaxies that have reported elevated FIR/radio ratios relative to the canonical correlation value for individual star-forming regions \citep[e.g.,][]{ejm06a,ah06,ejm08}.  
The physical explanation for this observation is that CR electrons propagate significantly further than dust-heating photons, thus the non-thermal radio continuum emission associated with individual star-forming regions is much more spread out and diffuse than the dust emission -- an effect which is not taken into account by the SFR recipe (see $\S$\ref{sec-sfrcr}).    

\subsection{Warm Dust: 24\,$\mu$m Emission}
Focusing on only the warm dust emission, we compare 24\,$\mu$m-derived SFRs against the 33\,GHz suite of SFRs in the third row of panels in Figure \ref{fig-3}.  
Unlike the total IR-based SFRs, there is much better agreement between the 24\,$\mu$m-derived and free-free emission SFRs among all extranuclear star-forming regions.  
There is also relatively good agreement among the 24\,$\mu$m-derived and total 33\,GHz SFRs among all extranuclear star-forming regions, while the 24\,$\mu$m-derived SFR for the nucleus is a factor of $\sim$2 larger than each of the 33\,GHz SFR estimates.   
Similar to the discrepancy between the IR-based and 33\,GHz free-free SFRs for the nucleus, we may expect the 24\,$\mu$m-derived and 33\,GHz free-free SFRs to also disagree because of either the capturing of ionizing photons by dust, or enhanced heating of very small grains (see $\S$\ref{sec-nucdiff}).  
Such grains are stochastically heated and only depend on the strength of the radiation field energy density.  
Warm dust may also arise from active galactic nuclei (AGN) activity, however there is currently no strong evidence for AGN activity in the nucleus of NGC\,6946 \citep[e.g.,][]{tsai06}.  
And, like other comparisons with the non-thermal SFRs, the 24\,$\mu$m derived SFRs are a factor of $\sim$3 larger than the non-thermal 33\,GHz SFRs, on average.  

\subsection{Extinction Corrected H$\alpha$: H$\alpha + 24\,\mu$m}
The hybrid H$\alpha + 24\,\mu$m SFR diagnostic has been found to be quite reliable relative to SFRs measured using extinction-corrected Pa$\alpha$ recombination line observations \citep[e.g., HST NICMOS narrowband imaging;][]{dc07}.  
In the fourth row of panels, we plot this SFR diagnostic against each of the 33\,GHz SFRs, keeping in mind that we believe that the free-free emission likely gives the most reliable SFR.  
We find a similar behavior here as with the SFRs derived using warm dust emission.  
The H$\alpha + 24\,\mu$m SFRs agree quite well with the 33\,GHz free-free SFRs for the extranuclear star forming regions, however, it is slightly larger than (i.e., 60\% above) the 33\,GHz thermal estimate in the nucleus.  
We find the same behavior in the comparison between the H$\alpha + 24\,\mu$m and total 33\,GHz SFRs.  
The non-thermal 33\,GHz SFRs again underestimate the SFRs among the extranuclear star-forming regions relative to the H$\alpha + 24\,\mu$m SFRs by an average factor of $\sim$4.  

\subsection{Bolometric Estimates: UV + IR}
Finally, we compare our last extinction-insensitive SFR diagnostic (i.e., UV + IR SFRs) with the 33\,GHz SFRs in the bottom panels of Figure \ref{fig-3}.  
In general, we find that the UV + IR SFRs are in good agreement with the 33\,GHz free-free SFRs among the extranuclear star-forming regions.  
The case is marginally improved when comparing the  UV + IR SFRs to the total 33\,GHz SFRs.  
Not surprisingly, we again find that the 33\,GHz non-thermal SFRs are quite discrepant, begin a factor of nearly $\sim$3 lower than the UV + IR SFRs.  
And, in all 3 comparisons, we find that the nucleus UV + IR SFR is a factor of $\ga$2 larger than the 33\,GHz SFRs.

\section{Discussion}
In the previous section we have quantitatively compared a number of SFR diagnostics whose reliability are considered to be largely insensitive to interstellar extinction for 10 star-forming regions within the nearby galaxy, NGC\,6946.  
Using these results, we attempt to construct a self-consistent picture to explain the observed discrepancies between the SFR estimates for the nuclear and extranuclear regions by using free-free, dust, and non-thermal emission processes as diagnostics.    
Additionally, we discuss how combinations of these diagnostics can be used to infer other physical characteristics from star-forming regions, and star-forming galaxies as a whole.    

\begin{figure*}
\begin{center}
\scalebox{1.0}{
\plottwo{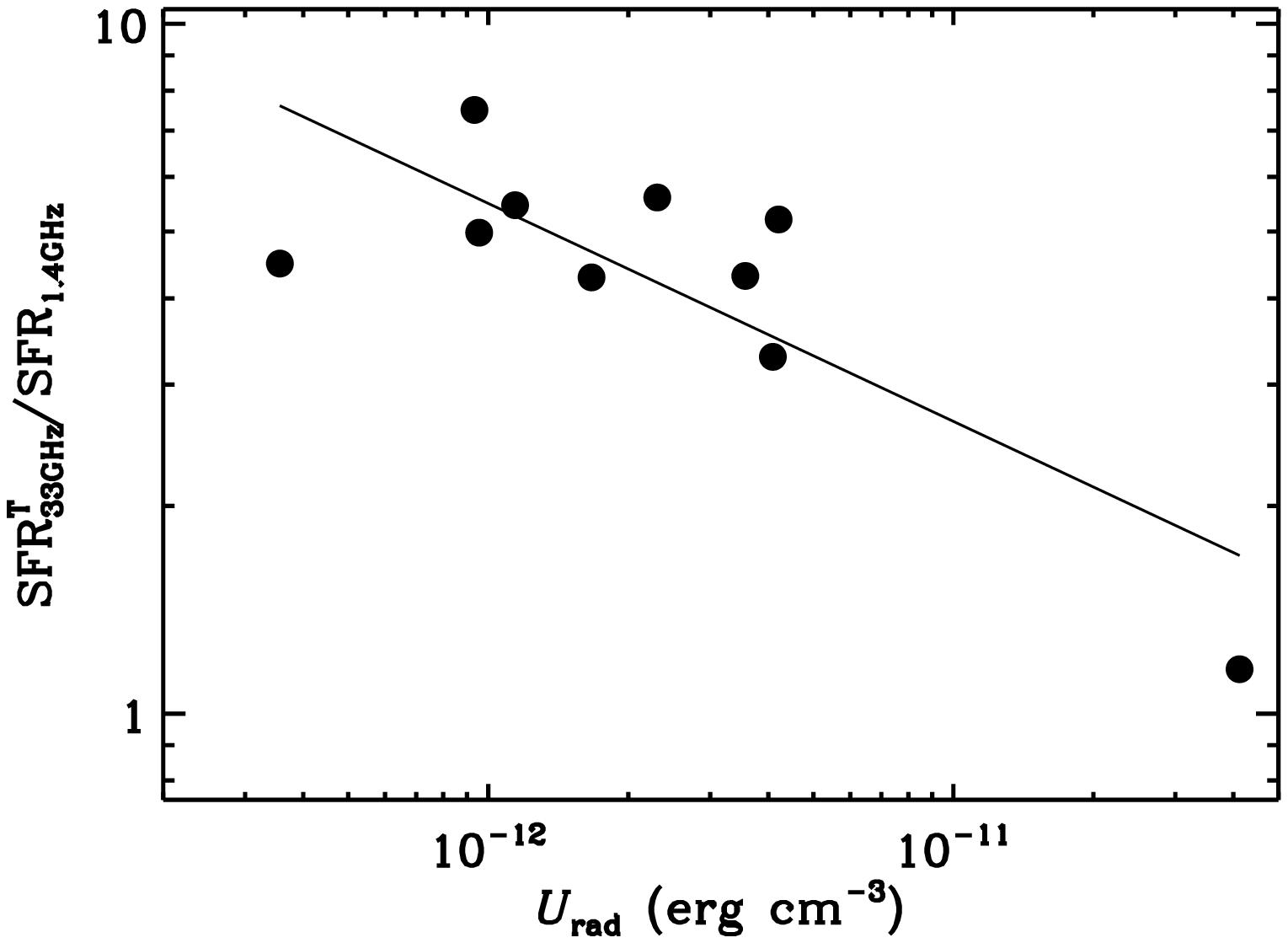}{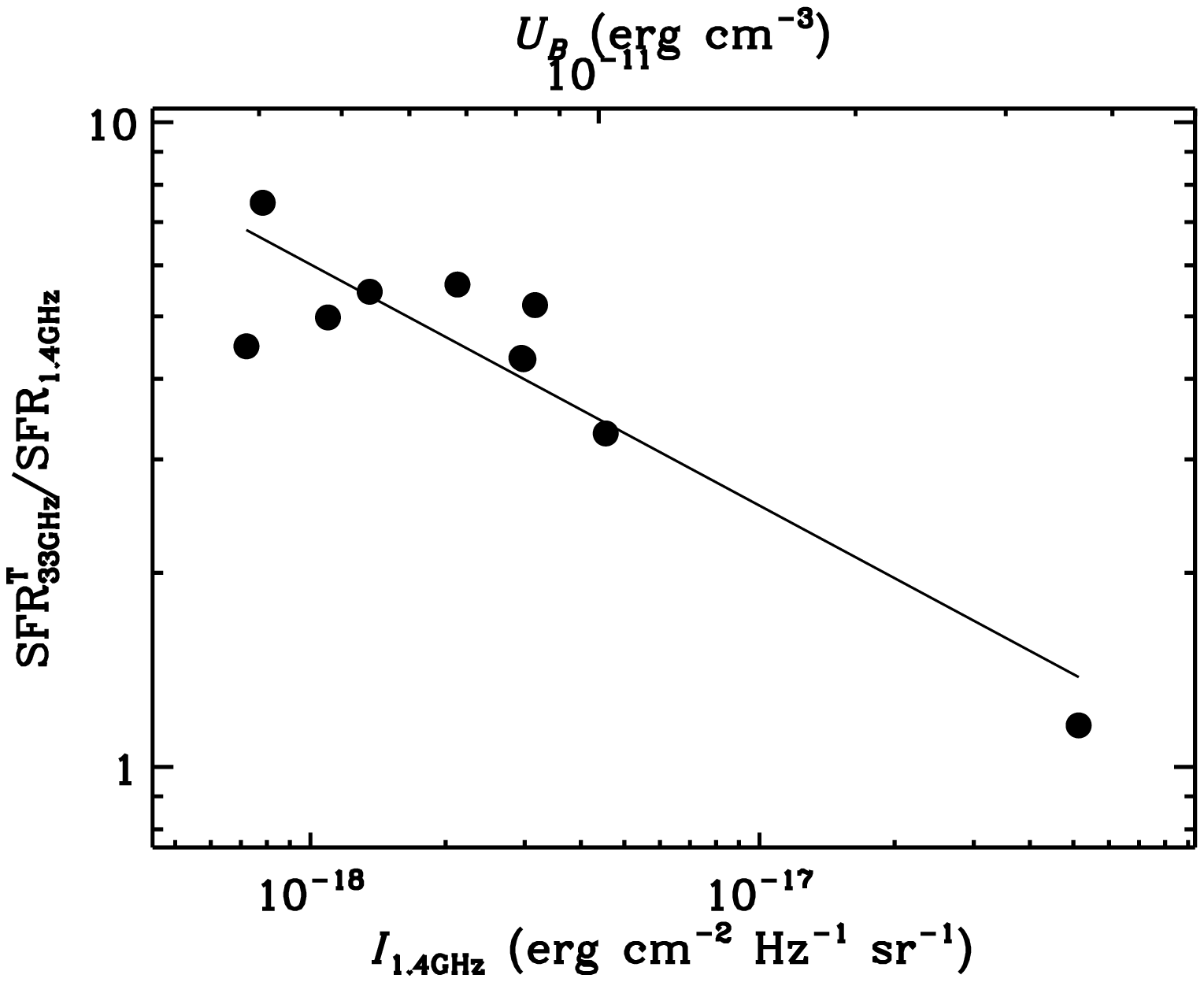}}
\caption{
The ratio of the 33\,GHz free-free 
to 1.4\,GHz (non-thermal) SFRs plotted against $U_{\rm rad}$ (left panel) and 1.4\,GHz surface brightness (right panel).  
Ordinary least squares fits to the trends 
are shown.  
\label{fig-4}}
\end{center}
\end{figure*}

\subsection{Anomalous Behavior of the Nuclear Starburst \label{sec-nucdiff}}
One of the most striking differences in the present comparison among these star-forming regions is the behavior between the nucleus and the extranuclear star-forming regions.  
While the 1.4\,GHz radio continuum emission and the FIR-radio correlation
yield a SFR that agrees well with the 33\,GHz free-free SFR for the nucleus, we find that all other diagnostics that rely on some sort of dust emission yield SFRs that are significantly larger than that when using the 33\,GHz free-free emission.    

As previously stated, one way to explain the dust-based SFRs being larger than the 33\,GHz free-free SFR estimates in the nucleus is through dust capture of ionizing photons, which will in turn increase the dust emission while decreasing the H$\alpha$ and free-free emission per unit star formation.  
The extinction in NGC\,6946's nuclear starburst  has been measured to be $A_{V}\sim4.3$~mag \citep[8\farcs5$\sim$280~pc;][]{ce96} and $A_{V}\sim4.6$~mag \citep[2\farcs7$\sim$90~pc;][]{qy01} using Balmer decrements.  
Similar values have been estimated using the 9.7\,$\mu$m silicate absorption feature in the central kpc \citep[$A_{V}\sim5.0$~mag;][]{jds07}, while much larger estimates have been quoted for the central $\sim$60~pc using total gas masses \citep[$A_{V}\sim100$~mag;][]{es06}, indicating a highly obscured, compact component.  
Although the extinction is high towards the central starburst, the effect of dust capturing ionizing photons may not significantly affect our measurements over the entire $\sim$0.8\,kpc area covered by the GBT beam unless the starburst is highly concentrated within the central $\sim$60~pc.  

Using our 24\,$\mu$m image at its native resolution, we find that $\sim$25\% of this warm dust emission arises within the central 200~pc.  
Assuming that 50 to 100\% of this emission from the central starburst region (i.e., 12.5 to 25\% of the total 24\,$\mu$m emission) is powered by ionizing photons, which are thus not participating in the production of free-free emission, this does not account for the factor of $\sim$2 discrepancy between the SFR estimates as the 24\,$\mu$m SFR would still be larger than the 33\,GHz free-free SFR by $\sim$55 to 20\%, respectively.      
This estimate relies on the relation between the 24\,$\mu$m and free-free emission given by Equation~2 in \citet{ejm06b}.  
It is also worth pointing out that the non-thermal (e.g., 1.4\,GHz) emission should not be affected by this competition for ionizing photons and that these SFRs (i.e., the 1.4\,GHz and 33\,GHz non-thermal) agree with the 33\,GHz free-free SFR.  
Thus, it seems that dust absorption of ionizing photons by itself cannot explain this discrepancy.  

Alternatively, it may be more likely that these relatively larger dust-based SFRs are the result of extra dust heating in the nucleus not associated with current star formation activity due to its large stellar density.  
In this case, the dust-based SFR will overestimate the true SFR for the nucleus of NGC\,6946.  
Such a scenario seems plausible given the large stellar densities associated with the pseudobulge of NGC\,6946 \citep{korm10} which completely fills our beam; 
the radius of the bulge is $\approx$15\arcsec\ \citep{rv95} and contains a stellar minibar \citep{dme98}.  
Furthermore, excess heating of dust (small and large grains) may arise from an accumulation of non-ionizing stars in the nucleus as the result of an extended duration of star formation activity associated with the starburst.    
Spectroscopic studies of near-infrared recombination lines have found that star formation episodes in nuclei differ significantly from those in extranuclear H{\sc ii} regions.  
Unlike extranuclear star-forming regions, galaxy nuclei typically show a spatial coincidence between the continuum and line emission suggesting that nuclear star formation activity is typically ongoing,  where as the star-forming episodes in extranuclear regions are much younger, and more transient in nature \citep[][]{dd04}.  

\subsubsection{Galaxy Nuclei Behaving More Like Entire Galaxies}
Assuming that ionizing photons are not being captured by dust, it is interesting to point out that if we instead extinction correct the H$\alpha$ line emission using the relation given by \citet{rck09} for entire galaxies, which scales the 24\,$\mu$m emission by a lower value (0.020 versus 0.031 for H{\sc ii} regions), we find good agreement between the H$\alpha + 24\,\mu$m and 33\,GHz free-free SFRs (crosses plotted in Figure \ref{fig-3}).  
Thus, it appears that the nucleus may behave more like an entire galaxy with respect to the non-thermal radio and H$\alpha+24\mu$m SFR calibrations.   
Assuming then, that our UV + IR SFR calibration is only appropriate for extranuclear star-forming complexes, we can set our 33\,GHz free-free SFR equal to Equation \ref{eq-sfrtot} to derive a new scaling to correct down the UV + IR SFR for the nucleus such that, 
\begin{equation}
\label{eq-sfrtotnuc}
\left(\frac{\rm SFR_{tot}^{nuc}}{M_{\sun}~{\rm yr^{-1}}}\right) = 4.42\times10^{-44} \left[\frac{L_{\rm FUV} + 0.43 L_{\rm IR}} {\rm erg~s^{-1}} \right].  
\end{equation}
The coefficient scaling the IR emission is actually more in line with a detailed study of UV and IR emission from global measurements of normal star-forming galaxies, which report a value of $\sim$0.46 (C. Hao, et al. 2011, in preparation).  

Admittedly, the fact that the FIR-radio correlation appears to hold for the nucleus, and the non-thermal radio (i.e., 1.4\,GHz and 33\,GHz non-thermal) SFRs agree with the 33\,GHz free-free SFR here, could be fortuitous.  
Ultimately, it seems impossible to completely disentangle these two scenarios as there is no alternative way to accurately measure the ionizing photon rate.  
Thus, for such highly obscured regions, it appears difficult to accurately measure SFRs to better than a factor of $\sim$2.  


\subsection{Non-thermal Emission, SFRs, and CRs \label{sec-sfrcr}}
\label{sec-crsfr}
The fact that the prescriptions for converting the non-thermal emission into a SFR are doing a reasonable job 
for the nucleus relative to the extranuclear star-forming regions may be the result of the time dependence on CR electron diffusion.  
Work on the propagation of CR electrons in external galaxies has found a trend of decreasing CR electron propagation length with increasing radiation field energy density, as measured by the infrared surface brightness \citep{ejm06b,ejm08}.  
The radiation field energy density in the nucleus, which appears to be in rough equipartition with the magnetic field energy density, is a factor of nearly $\sim$25 times larger than that of the extranuclear star-forming regions, on average.  
This is consistent with a scenario of underestimated 1.4\,GHz SFRs in the extranuclear star-forming regions relative to the nucleus.  
We can see if this scenario is valid by comparing the cooling efficiencies and diffusion timescales for the nucleus and extranuclear star-forming regions.  

Let us assume that the propagation of CR electrons is described by a random walk with an energy-dependent diffusion coefficient $D_{E}$ \citep[e.g.,][]{ginz80}, where $D_{E} = 5\times10^{28}~{\rm cm^{2}~s^{-1}}$ for $E < 1$~GeV and $5\times10^{28}(E/{\rm GeV})^{1/2}~{\rm cm^{2}~s^{-1}}$ for $E \geq 1$~GeV \citep[e.g.,][]{fj01,im02,dm02}.  
Then, the time it takes to diffuse a distance $l_{\rm diff}$ is given by $\tau_{\rm diff} = l_{\rm diff}^{2}/D_{E}$.  
For 1.4\,GHz emitting CR electrons to travel a distance of $\approx$0.8\,kpc (i.e., to cross the GBT beam at 33\,GHz), it will take $\sim$3.0 and 2.4~Myr in the nucleus and extranuclear regions, on average, respectively.  
The slight difference in $\tau_{\rm diff}$ arises from differences in the typical energy of 1.4\,GHz emitting electrons in the nucleus and extranuclear regions due to differences in the estimated magnetic field strengths.    
Accordingly, CR electrons in extranuclear star-forming regions will rapidly propagate away from their acceleration sites on timescales that are much longer than (i.e., a factor of $\sim$5) estimates of their typical cooling times ($\sim$12~Myr). 
On the other hand, the estimated cooling time for 1.4\,GHz emitting CR electrons in the nucleus is $\sim$2.0~Myr, which is $\sim$1.0~Myr shorter than the estimated escape time.  

Thus, the agreement between between the non-thermal and 33\,GHz free-free SFRs for the nucleus likely arises from being in the calorimeter limit of \citet{hv89} such that the continuously injected CR electrons, associated with the extended duration of star formation in the nucleus as discussed in $\S$\ref{sec-nucdiff}, lose the bulk of their energy before propagating outside of the region covered by our beam.  
It is also worth noting that the large gas densities in the nucleus will lead to an increase in hadronic interactions between CR nuclei and the interstellar gas, netting more synchrotron radiation per unit star formation due to an increase in pionic secondaries ($e^{\pm}$).  
On the other hand, the transient nature of the star formation episodes in the young extranuclear star-forming complexes provides enough time for CR electrons to propagate significantly further (i.e., on the order of a $\sim$kpc) than dust-heating and ionizing photons, resulting in an underestimate of the true SFR.

\subsubsection{Revised Non-Thermal SFR Recipes for Individual Star-Forming Regions}
It may be possible to correct the non-thermal SFR recipes by including a term that takes into account the time dependence of CR propagation.  
For example, through a detailed comparison between the non-thermal radio and infrared  morphologies of nearby disk galaxies, 
it has been shown that $U_{\rm rad}$ can be used as a proxy for the average distance traveled by the CR electrons \citep{ejm06b,ejm08}.  
We therefore plot the ratio of the 33\,GHz free-free to 1.4\,GHz (non-thermal) SFRs against $U_{\rm rad}$ (left panel of Figure \ref{fig-4}).  
We apply an ordinary least squares fit to derive a rough correction for the two SFR recipes that rely on the non-thermal radio emission (i.e., Equations \ref{eq-sfrnt} and \ref{eq-sfrq}).  
Accordingly, we can re-write Equation \ref{eq-sfrnt} such that
\begin{equation}
\label{eq-sfrntcorr}
\begin{split}
\left(\frac{\rm SFR_{\nu}^{NT-corr}}{M_{\sun}~{\rm yr^{-1}}}\right) &\sim 5.89^{+85}_{-5.5}\times10^{-32}\\
&  \left(\frac{\nu}{\rm GHz}\right)^{\alpha^{\rm NT}}
	\left(\frac{L_{\nu}^{\rm NT}}{\rm erg~s^{-1}~Hz^{-1}}\right)\\
&	\left(\frac{U_{\rm rad}}{\rm erg~cm^{-3}}\right)^{-0.316\pm0.102}, 
\end{split}
\end{equation}
and Equation \ref{eq-sfrq} such that
\begin{equation}
\label{eq-sfrqcorr}
\begin{split}
\left(\frac{\rm SFR_{\rm 1.4GHz}^{\rm corr}}{M_{\sun}~{\rm yr^{-1}}}\right) &\sim 5.64^{+81}_{-5.3}\times10^{-32}\\
&\left(\frac{L_{\rm 1.4GHz}}{\rm erg~s^{-1}~Hz^{-1}}\right) 
	\left(\frac{U_{\rm rad}}{\rm erg~cm^{-3}}\right)^{-0.316\pm0.102}.   
\end{split}
\end{equation}
Uncertainties from the least squares fits are given.  
Using these revised expressions results in ${\rm <SFR_{33GHz}^{NT-corr}/SFR_{33GHz}^{T}>} \sim 1.19 \pm 0.36$ and ${<\rm SFR_{1.4GHz}^{corr}/SFR_{33GHz}^{T}>} \sim 1.07 \pm 0.33$.  

Since $U_{\rm rad}$ should roughly scale with $U_{B}$ \citep[e.g., given the tightness of FIR-radio correlation;][]{jc92}, and $U_{B}$ scales as the non-thermal radio surface brightness, it is likely more convenient to try to derive a correction which relies on the non-thermal radio emission alone.  
We therefore plot the ratio of the 33\,GHz free-free 
to 1.4\,GHz (non-thermal) SFRs against the 1.4\,GHz surface brightness in the right panel of Figure \ref{fig-4} and perform a least squares fit.  
This results in corrections to the SFR recipes relying on the non-thermal radio continuum where Equation \ref{eq-sfrnt} becomes 
\begin{equation}
\label{eq-sfrntcorr2}
\begin{split}
\left(\frac{\rm SFR_{\nu}^{NT-corr}}{M_{\sun}~{\rm yr^{-1}}}\right) &\sim 7.33^{+64}_{-6.6}\times10^{-35}\\
&  \left(\frac{\nu}{\rm GHz}\right)^{\alpha^{\rm NT}}
	\left(\frac{L_{\nu}^{\rm NT}}{\rm W~Hz^{-1}}\right)\\
	&\left(\frac{I_{\rm 1.4GHz}}{\rm erg~s^{-1}~cm^{-2}~Hz^{-1}~sr^{-1}}\right)^{-0.374\pm0.057}, 
\end{split}
\end{equation}
and Equation \ref{eq-sfrq} becomes 
\begin{equation}
\label{eq-sfrqcorr2}
\begin{split}
\left(\frac{\rm SFR_{\rm 1.4GHz}^{\rm corr}}{M_{\sun}~{\rm yr^{-1}}}\right) &\sim 7.01^{+61}_{-6.3}\times10^{-35}\\
&\left(\frac{L_{\rm 1.4GHz}}{\rm erg~s^{-1}~Hz^{-1}}\right)	\\
&	\left(\frac{I_{\rm 1.4GHz}}{\rm erg~s^{-1}~cm^{-2}~Hz^{-1}~sr^{-1}}\right)^{-0.374\pm0.057}.   
\end{split}
\end{equation}
These revised expressions result in ${\rm <SFR_{33GHz}^{NT-corr}/SFR_{33GHz}^{T}>} \sim 1.15 \pm 0.32$ and ${<\rm SFR_{1.4GHz}^{corr}/SFR_{33GHz}^{T}>} \sim 0.98 \pm 0.22$.  
Accordingly, these revised recipes may be more appropriate when estimating SFRs from individual star forming regions within galaxies that have had local background emission removed.  


\begin{figure}
\begin{center}
\scalebox{1.2}{
\plotone{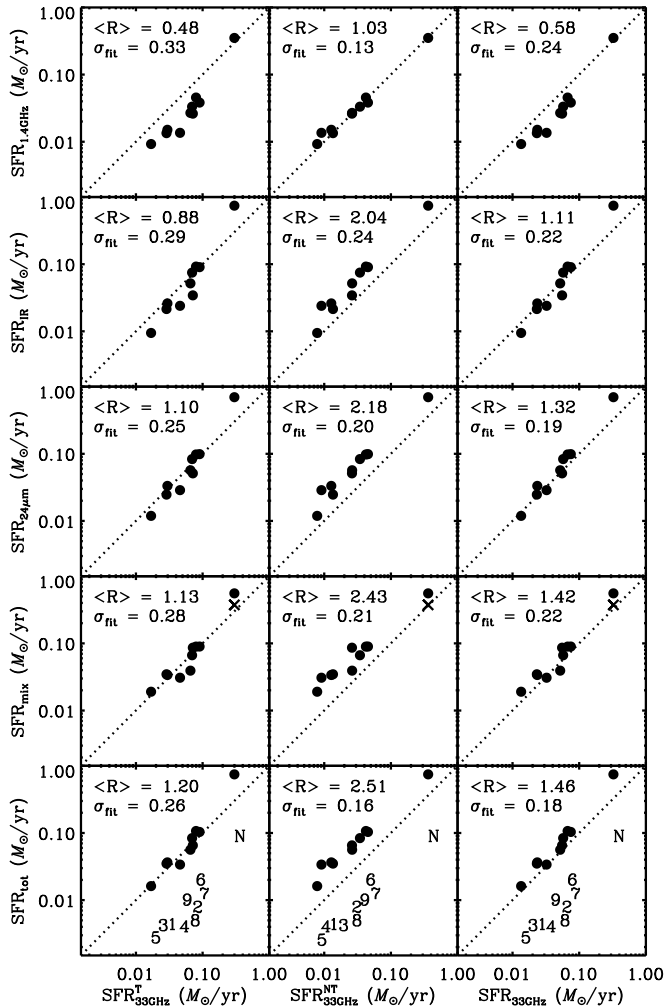}}
\caption{
The same as Figure \ref{fig-3} except that SFRs have been calculated using the photometry {\it without} local background subtracitons.  
\label{fig-5}}
\end{center}
\end{figure}

\subsection{Physical Parameters}
If we assume that Equation \ref{eq-hacorr} provides accurate (extinction-corrected) H$\alpha$ fluxes, and that the thermal radio continuum has been precisely quantified, we can compare these two quantities to obtain a rough idea of the electron temperature in each star-forming complex such that, 
\begin{equation}
\label{eq-te}
\begin{split}
\left( \frac{T_{\rm e}}{10^{4}~{\rm K}}\right) &\sim 3.68\times10^{23} \\
&\left[ 
\left(\frac{S_{\nu}^{\rm T}}{\rm erg~s^{-1}~cm^{2}~Hz^{-1}}\right)\right.\\
&\left.\left(\frac{f_{{\rm H}\alpha}^{\rm corr}}{\rm erg~s^{-1}~cm^{-2}}\right)^{-1}
\left(\frac{{\nu}}{\rm GHz}\right)^{0.1}
\right]^{1.695}
\end{split}
\end{equation}
\citep{cd86,jc92}.  
This relation assumes an ionized He to H density ratio of $N({\rm He}^{+})/N({\rm H}^{+})\sim0.08$.  
The approximation for the temperature dependence of recombination line strengths are good to $\sim$1\% over electron densities and temperatures ranging between $10^{2} \la n_{\rm e} \la 10^{4}~{\rm cm}^{-3}$ and $5000 \la T_{\rm e} \la 10000~{\rm K}$, respectively \citep{hs87,jc92}.  
Under these assumptions, the scatter measured by plotting SFR$_{\rm mix}$ versus SFR$_{\rm 33GHz}^{\rm T}$ in Figure \ref{fig-3} is dominated by variations in the electron temperatures from each region.  

The derived electron temperatures (see Table \ref{tbl-4}) range between $4000 \la T_{\rm e} \la 14000$~K, with a median of $\sim$11000~K, for all star-forming regions except for extranuclear regions 4 and 8 where unrealistically high temperatures are found, being 17000 and 28000~K, respectively.  
There are two possible reasons for such an outcome.  
One possibility is that the empirical H$\alpha$ extinction correction could be underestimating the true H$\alpha$ luminosity in these regions.    
This scenario seems plausible given that the extinction correction is built upon an empirical correlation which has scatter, and different calibration coefficients for different luminosity ranges.    

Alternatively, the thermal fraction from these regions may be overestimated.  
We note that the 33\,GHz flux density of extranuclear region 8 was boosted due to potential over-subtraction of background emission which may not have been adequate.  
For extranuclear region 4, the situation may be more complex as the 33\,GHz flux density from this source has been found to be a factor of $\sim$2 larger than what one would estimate by fitting the lower frequency (i.e., $\la$10\,GHz) radio data.  
This excess emission is thought to arise from rapidly spinning ultrasmall grains having a non-zero electric dipole moment.  
It is possible that the subtraction of this component was {\it underestimated}, however, the presence of such an excess indicates a high column of dust, which may suggest a possible underestimate in the H$\alpha$ extinction correction.  
It is also worth noting that an over prediction of the ionizing flux in extranuclear region 4 by the 33\,GHz free-free emission (i.e., after subtracting out the excess emission) relative to the H$\alpha+24\,\mu$m, 24\,$\mu$m, and IR observations (e.g., see Figure \ref{fig-4}) is opposite to the expectation if the excess emission was the result of optically-thick free-free emission arising from a compact H{\sc ii} region.  
Presently, it is not possible to distinguish between these two scenarios.  

It is also worth pointing out the low derived electron temperature for the Nucleus.   
Similar to the discussion in $\S$5.1, if we extinction correct the H$\alpha$ using the relation given by \citet{rck09} for entire galaxies, the electron temperature for the nucleus appears more reasonable, being $\sim$8000K.  

\subsection{Applicability to Global Measurements of Star-Forming Galaxies \label{sec-glob}}
While the current analysis has focused on individual star-forming complexes on $\sim$0.8\,kpc scales after removing a local diffuse background, it is instructive to investigate the significance of this subtraction in light of integrated measurements of star-forming galaxies for which a local subtraction is not possible.  
In Figure \ref{fig-5} we plot the same SFR comparisons as shown in Figure \ref{fig-3}, except that the IR and radio SED fitting was done using the photometry without subtracting out a local diffuse background component (see the Appendix for the associated fits and derived parameters).  
To briefly summarize the results from fitting the radio data in this manner, the observed radio spectral indices are steeper, being $\approx$0.5 on average, compared to $\approx$0.2 on average, when a local background was removed.    
The 33\,GHz emission is still clearly dominated by free-free emission as the thermal fraction at 33\,GHz is only slightly smaller when including the local background emission, being 79\%, on average.  

The trends found using these SFR estimates are nearly identical to those shown in Figure \ref{fig-3}.  
The SFRs estimated using the non-thermal emission are still significantly underestimated relative to the 33\,GHz free-free SFRs among the extranuclear regions, albeit by an average factor of $\sim$2 instead of $\sim$4$-$5.  
This reduction in discrepancy is certainly due to the fact that the bulk of the non-thermal radio emission is diffuse.  
It then follows that the corrections derived for the non-thermal SFR recipes in $\S$\ref{sec-sfrcr} will only be applicable to discrete star-forming regions for which a local background has been removed.  
Refitting these trends using the quantities derived before background subtraction leads to new coefficients for the SFR recipes in Equations \ref{eq-sfrntcorr}, \ref{eq-sfrqcorr}, \ref{eq-sfrntcorr2}, and \ref{eq-sfrqcorr2} of
of $5.35\times10^{-31}$, $5.11\times10^{-31}$, $3.31\times10^{-33}$, and $3.16\times10^{-33}$, respectively, while the exponents scaling $U_{\rm rad}$ and $I_{\rm 1.4GHz}$ become $-0.209$ and $-0.267$, respectively (see the Appendix for the re-written expressions).  
Interestingly, we again find good agreement between the non-thermal SFRs and the 33\,GHz free-free SFR for the nucleus.  

Among the SFRs which rely on some form of dust emission, we again find a reasonable agreement with the 33\,GHz free-free SFRs for the extranuclear star-forming regions.  
One notable difference is that the IR-based SFRs appear to be in much better agreement with the free-free emission SFRs by not having subtracted a local background. 
We also find that the nuclear SFRs are larger than the 33\,GHz free-free SFR for the nucleus by a factor of $\sim$2.5, on average.  
So, while the sample presented here has targeted bright, active star-forming regions, and is by no means representative of the global emission from a normal star-forming galaxy, the $\sim$0.8\,kpc scales investigated here do average over a number of discrete star formation sites which likely have a range in ages.    
Given this, and how well these diagnostics seem to agree even when including local diffuse emission, it appears that applying these diagnostics to entire systems which are vigorously forming stars is appropriate.  
In fact, it has recently been shown by \citet{dc10} that the H$\alpha +24\,\mu$m calibration used here, and derived for H{\sc ii} regions, works well for local IR-bright starbursts.  
Similarly, if we instead use the H$\alpha +24\,\mu$m calibration derived for entire galaxies \citep[i.e.,][]{rck09}, which has been found to be appropriate for the 24\,$\mu$m luminosities of each region \citep{dc10}, when local backgrounds are not removed, these SFRs agree extremely well with the 33\,GHz free-free SFRs (i.e., median ratio near unity) over the {\it entire} luminosity range investigated.  
Of course, applying such diagnostics to entire galaxies must be done with care.

\subsection{A Test for IMF Variations in High-$z$ Galaxies}
Recent work on low surface brightness dwarf galaxies have uncovered discrepancies between SFR diagnostics which depend on the ionizing photon rate (i.e., extinction corrected H$\alpha$) and the UV continuum \citep[e.g.,][]{gm09,jl09}.  
One scenario to explain these discrepancies is a change in the upper end of the IMF of these systems since the ionizing photon rate is much more sensitive to changes in the high-mass end of the IMF than the UV continuum.    
Following the same logic, it may be possible to measure such discrepancies, and identify variations in the upper end of the IMF, in high-$z$ objects using future high frequency ($\nu \ga 10$\,GHz) radio surveys, although it is worth noting that such variations can be mimicked by age effects, stochastic sampling of the IMF, and ionizing photon escape.  
The radio may be particularly powerful since, as we have shown, at higher frequencies the radio emission from galaxies becomes dominated by free-free emission.  
Surveys in the X-band ($8-12$\,GHz) can probe the rest-frame 30\,GHz emission from star-forming galaxies at $z\sim2$.  
In addition to measuring higher rest-frame frequencies with increasing redshift, the non-thermal component of a galaxy's radio continuum emission is expected to become increasingly suppressed with increasing redshift due to rapid cooling of CR electrons off of the cosmic microwave background \citep[][]{ejm09c}.    
Hints of such an effect have been observed \citep[e.g., flat radio spectral indices of lensed, high-{\it z}, galaxies;][]{rji10,si10}.  
This additional suppression of the non-thermal emission from galaxies could improve the situation by leaving only the thermal component detectable.  

\begin{figure}
\begin{center}
\scalebox{1.2}{
\plotone{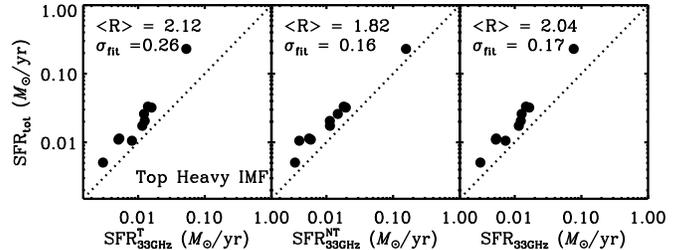}}
\caption{
The same as the bottom 3 panels in Figure \ref{fig-5}, except that a top-heavy Kroupa IMF was used, where the slope of the mass function is -1.5 for stellar masses ranging between $0.5-100~M_{\sun}$.
\label{fig-6}}
\end{center}
\end{figure}

In Figure \ref{fig-6} we present the same comparison of SFR diagnostics shown in the bottom 3 panels in Figure \ref{fig-5}, except that top-heavy IMF calibrations have been used. 
We again run Starburst99 under the same assumptions as described in $\S$\ref{sec-sfrcal}, except that the slope of the upper mass function has been flattened from $-2.3$ to $-1.5$.  
We choose a top heavy slope of $-1.5$ as this has been suggested as a way to explain the reionization of the intergalactic medium at $z \la 11$ by star-forming galaxies \citep{rrc08}.  
In the associated output the ionizing flux increases by a factor of $\approx$6 by moving to a top-heavy IMF, while the UV and IR emission only increases by a factor of $\approx$3.  
One of course expects that SFRs which depend only on the ionizing flux should be similarly independent of the IMF.  
However, based on these differences in the calibrations, SFRs proportional to the ionizing flux should be a factor of $\sim$2 times larger than those which depend on the UV and/or IR luminosities for a top heavy IMF compared to a standard Kroupa IMF.
In looking at Figure \ref{fig-6} this behavior is seen; 
the UV + IR SFRs yield a SFR which is nearly a factor of $\sim$2 discrepant from the 33 GHz thermal radio continuum estimate (i.e., a factor of $\sim$2 larger since the IMF of NGC\,6946 is more like a standard Kroupa IMF).   
Similarly, a systematic increase of $\sim$40\% is found when comparing the IR+UV SFRs to the SFRs estimated from the total 33\,GHz radio continuum.  

We note that in this last comparison, all three quantities used to estimate the SFRs are as observed; local backgrounds were not subtracted, no extinction correction was necessary, nor does it depend on an empirical calibration.  
Thus, by having high-quality UV, IR, and thermal radio continuum emission in hand, one should be able to quantify such systematic discrepancies between SFR diagnostics, and search for sources with potentially top heavy IMFs.


\section{Conclusions}
In the present analysis we have investigated the applicability and reliability of a number of SFR diagnostics that are not hampered by extinction from dust.  
They include 1.4\,GHz (non-thermal) radio, IR, 24\,$\mu$m, H$\alpha+24\,\mu$m, and UV + IR observations.    
This was done under the assumption that our free-free emission measurements in the Ka-band (i.e., at 33\,GHz) provide highly accurate estimates for the ionizing flux of massive star forming regions.  
Although we only focus on 10 star-forming regions in a single galaxy (i.e., NGC\,6946), the nuclear star-forming region is classified as a starburst and these regions span nearly 2 orders of magnitude in IR luminosity.  
Our main conclusions can be summarized as follows.   

\begin{enumerate}
\item
Excluding the total IR and non-thermal radio emission, there is very good agreement between the various diagnostics and the SFRs derived from the 33\,GHz free-free emission among the extranuclear star-forming regions, independent of local background subtractions.  
While the IR-based SFRs typically underestimate the SFRs derived from the free-free emission by an average factor of $\sim$2 among the extranuclear star-forming regions, there is significantly better agreement between these two diagnostics when local backgrounds are not subtracted.  
These differences likely arise from the subtraction of a local background and the application of a SFR calibration that assumes all UV photons are absorbed by dust at the source.



\item
Each of the diagnostics that rely on some sort of dust emission component yield SFRs for the nucleus that are larger than the 33\,GHz free-free emission SFR by an average factor of $\sim$2.  
Given that the 1.4\,GHz SFR agrees with the 33\,GHz free-free emission SFR, and that only $\sim$25\% of the 24\,$\mu$m emission within our beam arises from the central starburst, it seems most likely that these discrepancies are the result of excess IR heating arising from an accumulation of non-ionizing stars associated and extended duration of star formation in the nucleus rather than the capturing of ionizing photons by dust.    
However, conclusively distinguishing between these two scenarios is difficult.  

\item 
Typical SFR recipes that rely on non-thermal radio continuum emission (e.g., 1.4\,GHz SFRs based on the FIR-radio correlation) underestimate the true SFRs by an average factor of $\sim$4$-$5, or an average factor of $\sim$2 when local backgrounds are not removed, among the extranuclear star-forming regions.   
This is in contrast with the nucleus, for which the non-thermal emission estimates agree rather well with the 33\,GHz free-free emission.  
We believe that the agreement between the non-thermal and 33\,GHz free-free SFRs arises from CR electrons decaying within the nuclear starburst region with negligible escape, 
mirroring its high optical depth to ionizing photons.  
On the other hand, the transient nature of star formation episodes in the young extranuclear star-forming complexes allows for CR electrons to diffuse significantly further than dust heating photons, resulting in an underestimate in the SFR using the non-thermal emission.  
To properly measure the SFRs from individual star-forming regions using non-thermal emission requires a revised SFR recipe that attempts to incorporate the time-dependence of CR propagation. 

\item
Using the {\it total} 33\,GHz flux densities (i.e., not corrected for the presence of non-thermal emission) 
yields SFRs that are in excellent agreement with those measured using only the 33\,GHz free-free emission and combination of H$\alpha + 24\,\mu$m data among the extranuclear star-forming complexes.   
This result arises from the fact that free-free emission dominates  the total flux density at 33\,GHz (i.e., $\approx$87\%, on average, among the star-forming regions investigated) and suggests that observations in the rest-frame Ka-band may provide accurate SFRs in galaxies without the need of ancillary radio data to account for the non-thermal component.  

\item
Even without correcting our photometry for local diffuse background emission, which is prominent for observations that are most sensitive to non-thermal radio or cold dust emission,  the 33\,GHz flux densities are still dominated by free-free emission with thermal fractions of $\approx$79\%, on average.  
Furthermore, comparisons between the various SFR diagnostics remain generally unaffected, suggesting that these recipes are applicable to entire galaxies which are dominated by ongoing star formation activity.

\item 
Using a combination of high frequency ($\nu \ga10$\,GHz) radio data along with rest-frame far-infrared and UV data, one can compare radio versus UV + IR SFR estimates which should become increasingly discrepant (i.e., a factor of $\sim$2) as the IMF shifts towards one that is top heavy.  
This test makes use of the fact that radio continuum emission becomes dominated by thermal (free-free) emission at higher frequencies and, in addition, that the non-thermal emission of high-$z$ galaxies should become increasingly suppressed due to increased IC losses of CR electrons off of the cosmic microwave background.  
Thus, deep surveys at frequencies $\ga 10$\,GHz (e.g., X-band; $8-10$\,GHz) should begin to detect the free-free emission from galaxies at $z\ga2$ directly.  
Such tests to identify variations in the IMF of high-$z$ systems should be possible using deep survey imaging from upcoming radio (EVLA, MeerKAT \& SKA/NAA), mm/IR (ALMA/CCAT), and optical/NIR (JWST) facilities.  

\end{enumerate}

\acknowledgements
We would like the thank the referee, T. Takeuchi, for useful suggestions that helped to improve the content and presentation of this paper.  
EJM would like to thank the Observatories of the Carnegie Institution of Washington, where the final version of this paper was written.  
The National Radio Astronomy Observatory is a facility of the National Science Foundation operated under cooperative agreement by Associated Universities, Inc.
We are grateful to the SINGS team for producing high quality data sets used in this study. 
This work is based in part on observations made with the SST, which is operated by the Jet Propulsion Laboratory, California Institute of Technology under a contract with NASA. 
{\it Herschel} is an ESA space observatory with science instruments provided by European-led Principal Investigator consortia and with important participation from NASA.

\appendix
\section*{Estimates of Local Backgrounds and their Effect in the SED Fitting}
In Table \ref{tbl-5} we provide the values used for the local background subtraction at each wavelength for all star-forming regions.  
We also give the average fractional contribution of the background emission to the total emission per each wavelength and region at the end of each row and column, respectively.  
These values provide an estimate of the typical amount of diffuse emission contributing to each wavelength and region.  
Not surprisingly, the low frequency (i.e., 1.4, 1.5, and 1.7\,GHz) radio observations, which are dominated by non-thermal emission, are comprised of $\sim$50\% of diffuse emission, on average.  
The fractional contribution of the background to the total measurements at 850\,$\mu$m is anomalously low, which is most likely due to the insensitivity of these data.  
Comparing the average diffuse emission fraction among each star-forming complex, we see that the nucleus has the lowest contribution,  being a factor of $\sim$2.5 smaller at 14\%, on average.  

\begin{figure}
\begin{center}
\scalebox{0.6}{
\plotone{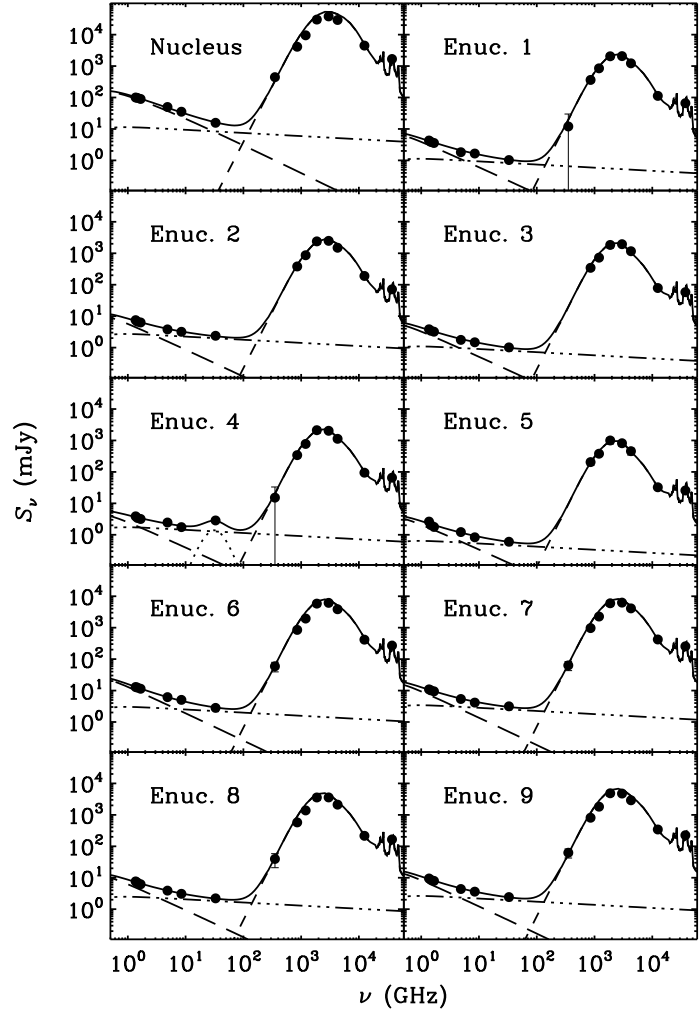}}
\caption{
The same as Figure \ref{fig-2}, except that the SEDs were fit using the photometry {\it without} subtracting out a local diffuse component.  
\label{fig-7}}
\end{center}
\end{figure}

Additionally, we show the modeled SEDs (Figure \ref{fig-7}) and give the associated derived parameters (Table \ref{tbl-6}) of each star forming region using the photometry before subtracting out a local diffuse component.  
The observed radio spectral indices are much steeper by not subtracting out a local diffuse component, being $\approx$0.5, on average, compared to $\approx$0.2, on average, when the backgrounds are removed.    
This finding is consistent with the low frequency non-thermal emission being dominated by a diffuse component.  
Also note worthy is the fact that the thermal fraction at 33\,GHz is still large, being $\approx$79\%, on average.  

\begin{deluxetable*}{lcccccccccc|c}
\tablecaption{Local Background Measurements \label{tbl-5}}
\tabletypesize{\scriptsize} 
\tablewidth{0pt}
\tablehead{
\colhead{Band}  &  
\colhead{Nuc.} & \colhead{Enuc. 1} &\colhead{Enuc. 2} &\colhead{Enuc. 3} &\colhead{Enuc. 4} &
\colhead{Enuc. 5} &\colhead{Enuc. 6} &\colhead{Enuc. 7}&\colhead{Enuc. 8} &\colhead{Enuc. 9}&
\colhead{$f_{\rm diff}^{\dagger}$\,(\%)}
}
\startdata

       $S_{\rm 1.4\,GHz}$~(mJy)&14.51&   2.98&   2.49&   2.00&   1.61&   1.43&   5.38&   5.63&   4.11&   4.56&  51\\
       $S_{\rm 1.5\,GHz}$~(mJy)&13.82&   2.79&   2.17&   1.78&   1.34&   1.01&   5.17&   5.10&   3.87&   4.14&  49\\
       $S_{\rm 1.7\,GHz}$~(mJy)&12.37&   2.49&   1.96&   1.56&   1.16&   0.81&   4.66&   4.60&   3.34&   3.74&  47\\
       $S_{\rm 4.9\,GHz}$~(mJy)& 5.14&   0.90&   0.71&   0.64&   0.65&   0.61&   1.90&   1.77&   1.49&   1.48&  33\\
       $S_{\rm 8.5\,GHz}$~(mJy)& 3.53&   0.80&   0.55&   0.55&   0.38&   0.26&   1.49&   1.23&   1.04&   1.13&  31\\
 $f_{\nu} (850\,\micron)$~(mJy)& 42.0&    1.2&    \nodata&    \nodata&    0.0&   \nodata&    4.3&    2.7&    0.0&    3.6&   5\\
  $f_{\nu} (350\,\micron)$~(Jy)& 0.99&   0.21&   0.16&   0.14&   0.17&   0.09&   0.48&   0.40&   0.31&   0.35&  45\\
  $f_{\nu} (250\,\micron)$~(Jy)& 2.71&   0.52&   0.38&   0.36&   0.43&   0.22&   1.26&   1.05&   0.83&   0.87&  55\\
  $f_{\nu} (160\,\micron)$~(Jy)& 6.48&   1.03&   0.74&   0.54&   0.63&   0.40&   2.81&   2.31&   1.66&   1.83&  38\\
  $f_{\nu} (100\,\micron)$~(Jy)& 5.90&   0.85&   0.54&   0.40&   0.39&   0.24&   2.38&   1.89&   1.32&   1.45&  30\\
   $f_{\nu} (70\,\micron)$~(Jy)& 3.12&   0.37&   0.22&   0.17&   0.13&   0.06&   1.23&   0.89&   0.57&   0.66&  21\\
  $f_{\nu} (24\,\micron)$~(mJy)&169.1&   24.7&   14.4&   12.1&   11.3&    7.8&   63.2&   55.0&   37.2&   41.9&  15\\
   $f_{\nu} (8\,\micron)$~(mJy)&170.6&   32.4&   18.3&   17.2&   13.4&   13.2&   77.2&   68.3&   50.8&   54.8&  28\\
       $f_{\rm H\alpha}$~($10^{-12}$\,erg\,s$^{-1}$\,cm$^{-2}$)& 0.34&   0.24&   0.16&   0.07&   0.07&   0.07&   0.32&   0.21&   0.13&   0.15&  16\\
    $f_{\nu} (1528$\,\AA)~(mJy)& 0.49&   0.93&   0.69&   0.31&   0.41&   0.36&   1.30&   0.84&   0.71&   0.69&  37\\
    $f_{\nu} (2271$\,\AA)~(mJy)& 1.74&   1.88&   1.67&   0.62&   0.94&   0.83&   2.86&   2.01&   1.82&   1.67&  40\\
\hline
          $f_{\rm diff}^{\dagger}$(\%)&14&  49&  21&  29&  22&  39&  40&  32&  46&  37&
\enddata
\tablecomments{$^{\dagger}$~The average fractional contribution of the
  background emission at each waveband and region (i.e., the fraction
  of diffuse emission) are given at the
  end of each row and column, respectively.}
\end{deluxetable*}

\begin{deluxetable*}{ccccccc}
\tablecaption{Derived Parameters Calculated {\it without} Local Backgrounds Subtracted \label{tbl-6}}
\tabletypesize{\scriptsize} 
\tablewidth{0pt}
\tablehead{
\colhead{ID}  & 
\colhead{$f_{\rm T}^{\rm 33\,GHz}$} & 
\colhead{$\alpha^{\rm obs \dagger}$} & \colhead{$\alpha^{\rm NT \dagger}$} &
\colhead{$L_{\rm IR}$}& \colhead{$B_{\rm min}^{\ddagger}$} & \colhead{$T_{\rm e}$}\\  
\colhead{} &
\colhead{(\%)} & \colhead{} &\colhead{} & \colhead{($10^{8} L_{\sun}$)}&
\colhead{($\mu$G)} & \colhead{($10^{4}$\,K)}
}
\startdata
     Nucleus   & 53$\pm$ 6  &0.62$\pm$0.04  &0.74$\pm$0.05  &50.10$\pm$ 2.43& 31.6$\pm$  9.7& 0.31$\pm$ 0.07\\
     Enuc. 1   & 80$\pm$ 5  &0.51$\pm$0.04  &0.81$\pm$0.06  & 1.76$\pm$ 0.11& 13.8$\pm$  4.0& 0.72$\pm$ 0.21\\
     Enuc. 2   & 83$\pm$ 6  &0.43$\pm$0.03  &0.83$\pm$0.06  & 2.31$\pm$ 0.14& 15.9$\pm$  4.6& 0.65$\pm$ 0.24\\
     Enuc. 3   & 79$\pm$ 7  &0.48$\pm$0.04  &0.81$\pm$0.06  & 1.45$\pm$ 0.10& 13.4$\pm$  3.9& 0.65$\pm$ 0.25\\
     Enuc. 4   & 45$^{*}\pm$ 5  &0.36$\pm$0.02  &0.81$\pm$0.06  & 1.62$\pm$ 0.10& 13.4$\pm$  3.9& 1.75$\pm$ 0.61\\
     Enuc. 5   & 78$\pm$ 8  &0.49$\pm$0.04  &0.80$\pm$0.06  & 0.64$\pm$ 0.06& 12.1$\pm$  3.5& 0.73$\pm$ 0.32\\
     Enuc. 6   & 79$\pm$ 6  &0.56$\pm$0.04  &0.84$\pm$0.06  & 6.21$\pm$ 0.30& 18.4$\pm$  5.3& 0.73$\pm$ 0.19\\
     Enuc. 7   & 79$\pm$ 6  &0.48$\pm$0.04  &0.84$\pm$0.06  & 6.08$\pm$ 0.31& 17.6$\pm$  5.0& 0.90$\pm$ 0.22\\
     Enuc. 8   & 82$\pm$ 6  &0.46$\pm$0.03  &0.83$\pm$0.06  & 3.49$\pm$ 0.18& 16.0$\pm$  4.6& 2.14$\pm$ 0.48\\
     Enuc. 9   & 79$\pm$ 6  &0.50$\pm$0.04  &0.84$\pm$0.06  & 5.05$\pm$ 0.26& 17.0$\pm$  4.9& 0.96$\pm$ 0.23
 \enddata
\tablecomments{$^{\dagger}$ The observed and non-thermal radio
  spectral indices measured between 1.4 and 8.5\,GHz.   
$^{\ddagger}$ Calculated using the total 1.4\,GHz flux
  densities and the revised minimum energy calculation of \citet{bk05}.    
$^{*}$ Calculated using the observed 33\,GHz flux density before subtracting out the $\approx$50\% contribution thought to arise from anomalous dust emission (see $\S$\ref{sec-modspec}). }
\end{deluxetable*}

As described in $\S$\ref{sec-glob}, the corrections derived for the non-thermal SFR recipes in $\S$\ref{sec-sfrcr}, which attempt to take into account the time dependence of CR propagation, will only be applicable to discrete star-forming regions for which a local background has been removed.  
By refitting these trends using the photometry without subtracting local backgrounds leads to different revised SFR recipes such that Equations \ref{eq-sfrntcorr} and \ref{eq-sfrqcorr} become
\begin{equation}
\label{eq-sfrntcorr3}
\left(\frac{\rm SFR_{\nu}^{NT-corr}}{M_{\sun}~{\rm yr^{-1}}}\right)  \sim 5.35^{+58}_{-4.9}\times10^{-31} \left(\frac{\nu}{\rm GHz}\right)^{\alpha^{\rm NT}}
	\left(\frac{L_{\nu}^{\rm NT}}{\rm erg~s^{-1}~Hz^{-1}}\right) \left(\frac{U_{\rm rad}}{\rm erg~cm^{-3}}\right)^{-0.209\pm0.094} 
\end{equation}
and 
\begin{equation}
\label{eq-sfrqcorr3}
\left(\frac{\rm SFR_{\rm 1.4GHz}^{\rm corr}}{M_{\sun}~{\rm yr^{-1}}}\right)  \sim 5.11^{+55}_{-4.7}\times10^{-31} \left(\frac{L_{\rm 1.4GHz}}{\rm erg~s^{-1}~Hz^{-1}}\right) \left(\frac{U_{\rm rad}}{\rm erg~cm^{-3}}\right)^{-0.209\pm0.094},    
\end{equation}
respectively.  
Using these revised expressions results in ${\rm <SFR_{33GHz}^{NT-corr}/SFR_{33GHz}^{T}>} \sim 0.95 \pm 0.27$ and ${<\rm SFR_{1.4GHz}^{corr}/SFR_{33GHz}^{T}>} \sim 1.03 \pm 0.29$.  
Similarly, by using the photometry before local background subtraction, Equations \ref{eq-sfrntcorr2} and \ref{eq-sfrqcorr2} become 
\begin{equation}
\label{eq-sfrntcorr4}
\left(\frac{\rm SFR_{\nu}^{NT-corr}}{M_{\sun}~{\rm yr^{-1}}}\right)  \sim 3.31^{+75}_{-3.2}\times10^{-33} \left(\frac{\nu}{\rm GHz}\right)^{\alpha^{\rm NT}}
	\left(\frac{L_{\nu}^{\rm NT}}{\rm W~Hz^{-1}}\right) \left(\frac{I_{\rm 1.4GHz}}{\rm erg~s^{-1}~cm^{-2}~Hz^{-1}~sr^{-1}}\right)^{-0.267\pm0.080}  
\end{equation}
and 
\begin{equation}
\label{eq-sfrqcorr4}
\left(\frac{\rm SFR_{\rm 1.4GHz}^{\rm corr}}{M_{\sun}~{\rm yr^{-1}}}\right)  \sim 3.16^{+72}_{-3.0}\times10^{-33} \left(\frac{L_{\rm 1.4GHz}}{\rm erg~s^{-1}~Hz^{-1}}\right) \left(\frac{I_{\rm 1.4GHz}}{\rm erg~s^{-1}~cm^{-2}~Hz^{-1}~sr^{-1}}\right)^{-0.267\pm0.080},    
\end{equation}
respectively.  These revised expressions result in ${\rm <SFR_{33GHz}^{NT-corr}/SFR_{33GHz}^{T}>} \sim 0.97 \pm 0.24$ and ${<\rm SFR_{1.4GHz}^{corr}/SFR_{33GHz}^{T}>} \sim 1.03 \pm 0.25$.  
Accordingly, these revised recipes may be more appropriate when estimating SFRs from individual star forming regions within galaxies that have not had local backgrounds removed.



\bibliography{/Users/emurphy/libs/bibtexref/master_ref}

\end{document}